\documentclass[manuscript]{acmart}
\usepackage{tikz}
\usepackage{listings}
\usepackage{tcolorbox}

\definecolor{codegreen}{rgb}{0,0.6,0}
\definecolor{codegray}{rgb}{0.5,0.5,0.5}
\definecolor{codepurple}{rgb}{0.58,0,0.82}
\definecolor{backcolour}{rgb}{0.95,0.95,0.92}

\lstdefinestyle{mystyle}{
    backgroundcolor=\color{backcolour},   
    commentstyle=\color{codegreen},
    keywordstyle=\color{magenta},
    numberstyle=\tiny\color{codegray},
    stringstyle=\color{codepurple},
    basicstyle=\ttfamily\footnotesize,
    breakatwhitespace=false,         
    breaklines=true,                 
    captionpos=b,                    
    keepspaces=true,                 
    numbers=left,                    
    numbersep=5pt,                  
    showspaces=false,                
    showstringspaces=false,
    showtabs=false,                  
    tabsize=2
}
\usepackage{xcolor}  

\AtBeginDocument{%
  }

\setcopyright{acmlicensed}
\copyrightyear{2025}
\acmYear{2025}
\acmDOI{XXXXXXX.XXXXXXX}

\acmJournal{TOSEM}

\begin{document}

\title{SDLog: A Deep Learning Framework for Detecting Sensitive Information in Software Logs}

\author{Roozbeh Aghili}
\email{roozbeh.aghili@polymtl.ca}
\orcid{0000-0002-9361-2369}
\affiliation{%
  \institution{Polytechnique Montreal}
  \city{Montreal}
  \country{Canada}
}

\author{Xingfang Wu}
\email{xingfang.wu@polymtl.ca}
\orcid{0000-0001-7040-3751}
\affiliation{%
  \institution{Polytechnique Montreal}
  \city{Montreal}
  \country{Canada}
}

\author{Foutse Khomh}
\email{foutse.khomh@polymtl.ca}
\orcid{0000-0002-5704-4173}
\affiliation{%
  \institution{Polytechnique Montreal}
  \city{Montreal}
  \country{Canada}
}

\author{Heng Li}
\email{heng.li@polymtl.ca}
\orcid{0000-0001-5441-6763}
\affiliation{%
  \institution{Polytechnique Montreal}
  \city{Montreal}
  \country{Canada}
}

\acmArticleType{Review}

\begin{abstract}
Software logs are messages recorded during the execution of a software system that provide crucial run-time information about events and activities. Although software logs have a critical role in software maintenance and operation tasks, publicly accessible log datasets remain limited, hindering advance in log analysis research and practices. The presence of sensitive information, particularly \textit{Personally Identifiable Information (PII)} and \textit{quasi-identifiers}, introduces serious privacy and re-identification risks, discouraging the publishing and sharing of real-world logs. In practice, log anonymization techniques primarily rely on regular expression patterns, which involve manually crafting rules to identify and replace sensitive information. However, these regex-based approaches suffer from significant limitations, such as extensive manual efforts and poor generalizability across diverse log formats and datasets. To mitigate these limitations, we introduce SDLog, a deep learning-based framework designed to identify sensitive information in software logs. Our results show that SDLog overcomes regex limitations and outperforms the best-performing regex patterns in identifying sensitive information. With only 100 fine-tuning samples from the target dataset, SDLog can correctly identify 99.5\% of sensitive attributes and achieves an F1-score of 98.4\%. To the best of our knowledge, this is the first deep learning alternative to regex-based methods in software log anonymization.
\end{abstract}

\keywords{Software logs, Data privacy, Anonymization, Regular expressions, Deep learning}

\ccsdesc[500]{Security and privacy~Data anonymization and sanitization}
\ccsdesc[500]{Computing methodologies~Neural networks}

\maketitle

\section{Introduction}
\label{sec:introduction}

Software logs are messages recorded during the execution of a software system, providing crucial run-time information about events and activities. Typically created through logging commands such as~\textit{logger.info()}, log message can include different components depending on the logging configuration. Figure~\ref{fig:log_example} shows a log statement generated using Python’s~\textit{logging} module and contains four components: a timestamp indicating when the event occurred (e.g., 2025-03-26 08:53:29,653), a log level representing the message's severity (e.g., INFO), the function name that produced the log (e.g.,~\textit{build\_log\_url}), and the message content, which forms the unstructured part of the log. Software logs are essential for developers, operators, and analysts to understand application execution~\cite{jiang2008abstracting}, diagnose system failures~\cite{nagaraj2012structured, cinque2012event}, detect anomalies~\cite{landauer2018dynamic, ryciak2022anomaly}, and monitor systems~\cite{barik2016bones, li2020qualitative}. In many real-world deployments, logs remain often the only resource for understanding and resolving complex system challenges~\cite{barik2016bones}. 

\begin{figure}[t]
  \centering
  \begin{minipage}{0.8\textwidth}
\begin{lstlisting}[
  language=Python,
  style=mystyle,
  frame=single,
  aboveskip=-0pt,
  belowskip=-0pt
]
def build_log_url(base_url, container_id, log_file):
    url = f"{base_url}/containerlogs/{container_id}/{log_file}"
    logging.info(f"The URL for getting the log: {url}")
    return url
\end{lstlisting}
  \end{minipage}

  \begin{center}
    \begin{tikzpicture}
      \node (arrow) {$\downarrow$};
    \end{tikzpicture}
  \end{center}

  \begin{minipage}{0.8\textwidth} 
\begin{lstlisting}[
  style=mystyle,
  frame=single,
  breaklines=true,
  breakindent=0pt,
  aboveskip=-0pt,
  belowskip=-0pt
]
2025-03-26 08:53:29,653 - INFO - build_log_url - The URL for getting the log: https://example.com/api/v1/containerlogs/container-123/application.log
\end{lstlisting}
  \end{minipage}
  \caption{A logging statement example in Python.}
  \label{fig:log_example}
\end{figure}

Modern software systems are complex and large-scale, generating huge volumes of logs (e.g., 120-200 million lines per hour~\cite{mi2013toward}). Despite this substantial data volume and the extensive research focused on various aspects of software logs~\cite{batoun2024literature, korzeniowski2022landscape}, publicly accessible log datasets remain limited. Several studies have highlighted this shortage and called for more open datasets~\cite{zhu2023loghub, aghili2023studying, bogatinovski2021artificial, aghili2024understanding}. Although initiatives such as LogHub~\cite{zhu2023loghub} and ITA~\cite{ita} have attempted to address this gap, privacy concerns continue to discourage organizations from sharing their production logs, making publicly available datasets scarce.

To share log datasets while protecting privacy, anonymization becomes a critical prerequisite. Privacy regulations such as the~\textit{General Data Protection Regulation (GDPR)} and~\textit{California Consumer Privacy Act (CCPA)} mandate strict protection of personal and potentially identifiable information~\cite{voigt2017eu}. Log entries frequently contain sensitive attributes such as user identifiers, IP addresses, host names, and system configurations, which could compromise individual or organizational privacy if exposed~\cite{aghili2025protecting}. Currently, anonymization techniques primarily rely on~ \textit{regular expression (regex)} patterns, which involve manually crafting rules to identify and replace sensitive information~\cite{varanda2021log, flegel2002pseudonymizing, lin2014pcaplib}. However, these regex-based approaches have significant limitations: they require manual effort, struggle with context-aware identification, and often fail to capture nuanced variations in log structures across different systems and logging frameworks. The static nature of regex makes them particularly ineffective for handling the complex, semi-structured text in modern software logs, where sensitive information can appear in diverse and unpredictable formats.

Given the limitations of regex-based anonymization, we introduce SDLog: \textbf{S}ensitivity \textbf{D}etector in Software \textbf{Log}s, a deep learning-based framework designed to identify sensitive information in software logs. SDLog builds upon the pre-trained~\textit{CodeBERT}~\cite{feng2020codebert}, which we fine-tune using sensitive attributes found in real-world logs. Our method uses CodeBERT’s contextual understanding to address the limitations of regex-based techniques, which rely on fixed patterns and often fail to generalize across diverse log structures. Unlike regex, SDLog dynamically interprets surrounding context, making it well-suited for the complex and unstructured nature of real-world software logs. We validate our approach by annotating and analyzing 32,000 log lines from different software systems, demonstrating high precision in identifying sensitive attributes such as IP addresses, user identifiers, and system paths. We evaluate the effectiveness of our approach by answering the following~\textit{Research Questions (RQs)}.

\begin{itemize}
\item [RQ1] \textbf{\textit{How effective are regex-based approaches in identifying sensitive attributes in software logs?}} We first investigate the effectiveness of regex-based methods in detecting sensitive attributes in software logs. To do this, we gather regex patterns from academic research, open-source projects, and industry practices, evaluating them on different sensitive attribute types. This evaluation allows us to identify variations, inconsistencies, and coverage gaps across different regex patterns.

\item [RQ2] \textbf{\textit{How effective is SDLog in identifying sensitive attributes in unseen software logs?}} We conduct a comprehensive evaluation across 16 diverse log datasets, benchmarking SDLog against the best set of regexes found in RQ1. Our evaluation consists of two parts: first, we assess SDLog's overall ability to distinguish sensitivity from non-sensitivity in software logs; second, we analyze SDLog's performance in accurately identifying specific categories of sensitive attributes, such as IP addresses, MAC addresses, and file paths.

\item [RQ3] \textbf{\textit{How effective is SDLog in identifying sensitive attributes when fine-tuned on a target dataset?}} In a real-world scenario, an organization might want to fine-tune SDLog using a subset of their log datasets to tailor the model’s sensitive attribute detection to their specific data characteristics. To measure the performance of SDLog in this scenario, we evaluate the fine-tuned SDLog using the same evaluation procedure as in RQ2, analyzing both the sensitive attribute detection of SDLog as well as its performance on the specific categorization of each sensitive attribute.
\end{itemize}

Our work makes several contributions:

\begin{enumerate}

\item We provide a systematic analysis of the effectiveness of regex-based approaches in detecting sensitive attributes in software logs.

\item We annotate a dataset of 32,000 log entries labeled with sensitive information categories.

\item We introduce SDLog, a novel method for identifying sensitive attributes in software logs.

\end{enumerate}

The rest of the paper is organized as follows. Section~\ref{sec:related_work} summarizes prior research related to our work. Section~\ref{sec:experiment} describes the experiment setup of our study, including the data preparation and the implementation details of SDLog. Section~\ref{sec:RQs} presents our methodology and results addressing the research questions. Section~\ref{sec:discussion} explores the implications of our findings. Section~\ref{sec:threats} discusses the threats to validity, and finally, Section~\ref{sec:conclusion} concludes our paper. We share our replication package information in Section~\ref{sec:data}.

\section{Background and Related Work}
\label{sec:related_work}
This work proposes a deep learning-based approach to identify sensitive attributes in software logs as a first step in an anonymization pipeline. Therefore, we discuss the background and related work in the following six subsections.

\subsection{PII and Sensitivity}
\label{sec:related_work_pii}
Preserving privacy in software logs is not just a technical challenge but a fundamental ethical and legal responsibility; it mitigates security risks, protects business confidentially, maintains user trust, ensures compliance with internal and external policies, and facilitates responsible data governance. A significant concern in software log management and sharing is the exposure of \textit{Personally Identifiable Information (PII)}. PII refers to any data that can be used to identify an individual, either directly, such as names and email addresses, or indirectly, such as device identifiers. While PII is often the focus of privacy concerns, it is not the only source of re-identification of individuals. A well-known study by Sweeney et al.~\cite{sweeney2002k} demonstrated that using just three attributes of postal code, date of birth, and sex, over 86\% of U.S. citizens could be uniquely identified. These attributes, known as \textit{quasi-identifiers}, are pieces of information that, while not directly revealing someone's identity on their own, can be combined with other data to potentially identify an individual.

Although data privacy has been extensively studied for decades, the particular challenges of preserving privacy in software logs have received limited attention. Prior research on log privacy and anonymization~\cite{gu2023pd, mcsherry2010differentially, xu2002prefix} often focus on a limited set of sensitive attributes, commonly treating only elements such as IP addresses as sensitive, while largely overlooking the role of quasi-identifiers within the context of software logs. In a recent attempt, Aghili et al.~\cite{aghili2025protecting} use a variety of approaches, such as literature review and survey, to conclude a selection of sensitive attributes in software logs. Their classification identifies IP addresses, MAC addresses, host names, file paths, IDs, URLs, usernames, port numbers, and configuration details as commonly sensitive attributes in software logs. In this study, we adopt this definition, attempting to identify these sensitive attributes in software logs using a machine learning approach.

\subsection{Privacy Regulations}
Privacy regulations are designed to protect personal data and ensure that they are handled responsibly and ethically. Many countries have introduced privacy regulations to guide how organizations collect, store, and use sensitive information. These regulations vary by region, and some countries enforce multiple laws to cover different types of data. In the United States, for example, the~\textit{Health Insurance Portability and Accountability Act (HIPAA)}~\cite{hipaa} addresses healthcare data, while the~\textit{California Consumer Privacy Act (CCPA)}~\cite{ccpa} focuses on consumer-related information. Similarly, China enforces two major laws:~\textit{Personal Information Protection Law (PIPL)}~\cite{pipl} and the~\textit{Data Security Law (DSL)}~\cite{dsl}. In Europe, the~\textit{General Data Protection Regulation (GDPR)}~\cite{gdpr} serves as the primary legal standard for data privacy. Likewise, Canada follows the~\textit{Personal Information Protection and Electronic Documents Act (PIPEDA)}~\cite{pipeda}, and Brazil has adopted the~\textit{General Data Protection Law (LGPD)}~\cite{lgpd}.

Several privacy regulations provide formal definitions of personal data and outline specific attributes that need protection. The GDPR, for example, defines personal data as any information that relates to an identified or identifiable individual. This broad definition covers both direct identifiers, such as names and ID numbers, and indirect identifiers, including IP addresses and location data. GDPR considers IP addresses as personal data when they can be linked to an individual, either directly or by combining them with other information accessible to the data controller. Unlike GDPR, which broadly defines personal data, HIPAA specifies 18 sensitive identifiers that must be protected. These include both direct identifiers (e.g., names, contact numbers, social security numbers) and indirect ones (e.g., geographic locations, date information, and IP addresses). Many of these identifiers are often present in software logs, such as timestamps, IP addresses, device or system identifiers, and URLs. While data protection laws, such as GDPR and HIPAA, provide guidance on what constitutes personal data, none are tailored specifically to the context of software logs. Hence, it is necessary to interpret and adapt relevant regulatory elements to log data, taking into account the applicable legal frameworks of each region.

\subsection{Sharing Software Logs and Traces}
Although publicly available software logs and traces are scarce, several initiatives have attempted to address this gap. One notable effort is LogHub~\cite{zhu2023loghub}, which is a collection of 16 datasets spanning six categories: distributed systems, supercomputers, operating systems, mobile systems, server applications, and standalone software. Widely adopted in the software engineering community, LogHub serves as a key dataset for research in several domains such as log parsing, anomaly detection, and system monitoring~\cite{balasubramanian2023transformer, pathak2024self}. However, despite its widespread usage, parts of the collection are considerably outdated; for instance, the~\textit{Apache} and~\textit{BGL} datasets date back to 2005. The~\textit{Internet Traffic Archive (ITA)}~\cite{ita} is another repository that offers server logs and packet traces. It includes large, real-world datasets such as~\textit{WorldCup98}, which records the 1.3 billion requests made to the servers of FIFA during the 1998 World Cup, or~\textit{Calgary}, which records the requests sent to the department of computer science of the University of Calgary in Canada over one year. However, all ITA datasets are relatively dated, having been collected between 1989 and 2007~\cite{aghili2024understanding}.

Certain cloud providers have also released datasets for public use. One example is the Google cluster workload dataset~\cite{verma2015large, google_cluster}, which contains traces collected from workloads managed by Google's cluster management system, Borg. This dataset has been released in three versions: the initial release in 2009, followed by updates in 2011 and 2019. The traces capture detailed information on job submissions, scheduling decisions, and resource usage associated with jobs executed within Borg clusters. Microsoft Azure has also released a dataset containing VM traces and Azure Functions traces~\cite{cortez2017resource, azure_cluster}. The VM traces are available in two versions, collected in 2017 and 2019. The 2017 dataset includes data from 2 million virtual machines and 1.2 billion resource utilization records, while the 2019 version has a record of 2.6 million VMs and 1.9 billion readings. Additionally, the Azure Functions traces, gathered in 2019 and 2021, provide insights into applications running on Azure Functions, capturing details such as application and function IDs, execution times, and memory usage. Since 2017, Alibaba has also released a series of cluster trace datasets derived from its production systems, including microservice-based architectures and GPU clusters~\cite{alibaba_cluster}. This repository is actively maintained, with the most recent update published in April 2025. The cloud traces from Google, Azure, and Alibaba are widely used across various research areas, including job scheduling, workload allocation, and failure prediction~\cite{li2019deepjs, sliwko2024cluster, tuns2023cloud}.

\subsection{Software Log Anonymization}
Despite the growing reliance on log data for debugging, monitoring, and analytics, log anonymization remains a challenging and underdeveloped task. While many attributes in software logs can reveal sensitive information, research in this area has predominantly concentrated on IP addresses as the only target for anonymization~\cite{aghili2025protecting}. One of the earliest and influential works on IP address anonymization is CryptoPAn~\cite{xu2002prefix}, introduced by Xu et al., which enables prefix-preserving anonymization of IP addresses. In this scheme, if two IP addresses share a k-bit prefix in their original form, their anonymized versions will retain that same prefix, preserving the hierarchical structure of the network. CryptoPAn applies a cryptographic bitwise process, where each anonymized bit is generated based on the bits previously anonymized. However, CryptoPAn is known to be vulnerable to fingerprinting and injection attacks~\cite{brekne2005anonymization, brekne2005circumventing}, in which adversaries employ known network flows or use static fields such as timestamps to inject synthetic flows in order to reverse-engineer anonymized addresses. Several subsequent tools, such as AFT-Anon~\cite{han2020aft}, PD-PAn~\cite{gu2023pd}, and the work by Manocchio et al.~\cite{manocchio2024configurable}, also focus solely on anonymizing IP addresses. Most of these tools adopt a prefix-preserving strategy, as removing the hierarchical structure of IP addresses can significantly reduce the utility of the data. To address the privacy limitations of prefix-preserving techniques, more recent methods have introduced additional anonymization layers. For instance, AFT-Anon~\cite{han2020aft} performs real-time anonymization using an in-memory~\textit{Active Flow Table (AFT)}, which maps original IPs to their anonymized versions. When a new packet arrives, it either retrieves the existing mapping or generates a new anonymized IP, updates the packet, and stores the mapping using a hash-based structure similar to a HashMap. In a different approach, Varanda et al.~\cite{varanda2021log} apply a~\textit{Hash-based Message Authentication Code (HMAC)} built on SHA256 to anonymize IP addresses, leveraging a secret key to enhance privacy and security beyond conventional hashing methods. Another example is the work by Xenakis et al.~\cite{xenakis2023self}, which integrates prefix preservation with clustering. They randomize the network prefix (usually the first three octets) through permutation and anonymize the host part (usually the last octet) by clustering similar addresses and replacing them with the cluster's average value across each 8-bit segment.

Although many research studies focus on IP address anonymization, other attributes, such as host names, file paths, and MAC addresses, receive very little attention~\cite{aghili2025protecting}. Only a limited number of studies address the anonymization of multiple attributes. One notable example is Flaim~\cite{slagell2006flaim}, which supports anonymization of a wide range of network-related fields, including IP addresses, MAC addresses, host names, port numbers, time stamps, and network protocols. For each of these attributes, Flaim offers a set of anonymization techniques. For example, it uses the black marker technique (i.e., replacing a value with a constant) to anonymize network protocols. However, a key limitation of Flaim is its restriction to network traffic data; it cannot be readily applied to other types of log datasets. Similarly, iCAT+~\cite{oqaily2023icat+} offers an anonymization space, where each attribute can be anonymized using a set of techniques. iCAT+ supports several attributes such as IP addresses, integers and decimal numbers, time stamps, and strings. It also incorporates a~\textit{Natural Language Processing (NLP)} model to translate the privacy requirements of data owners and data users. Several other tools also focus on anonymizing network data. TCPdpriv~\cite{tcpdpriv} is an anonymization tool designed for processing packet traces via the libpcap library, which is widely used in network traffic analysis. As a result, its functionality is restricted to network data. Additionally, its platform compatibility is limited, supporting only a few operating systems such as SunOS, Solaris, and FreeBSD. 
Expanding on the functionalities of TCPdpriv, IP2anonIP~\cite{ip2anonip} was developed to transform IP addresses into either host names or anonymized IPs. IP2anonIP also allows for the addition of custom fields. However, one major drawback is its inefficiency; processing a single day worth of data can take several hours. Other tools such as AnonTool~\cite{foukarakis2009deep, foukarakis2007flexible} and PktAnon~\cite{gamer2008pktanon} are tailored for packet-based data, particularly netFlow or libpcap formats. AnonTool, implemented in C, supports the anonymization of various attributes including IP addresses, timestamps, ports, response sizes, and netFlow-specific fields, such as~\textit{Type of Service (TOS)} and TCP flags. Likewise, PktAnon targets similar fields, focusing on anonymizing IP addresses and both TCP and UDP headers. In a similar design, AAPI~\cite{koukis2006generic} also works with network data and anonymizes attributes such as IP addresses, TCP and UDP fields, and URLs. All the mentioned frameworks either work only on network data, where attribute detection is unnecessary due to the structured format, or rely on regular expressions to identify attributes.

\subsection{Named Entity Recognition (NER)}
We model the detection of sensitive information in log messages as a \textit{Named Entity Recognition (NER)} task. Named Entity Recognition is one of the most common token classification tasks in NLP that attempts to find a label for each token in a natural language sequence~\cite{nadeau2007survey, li2020survey}. In NLP, NER serves not only as a standalone task for information extraction but also supports many downstream tasks, such as automatic text summarization~\cite{aone1999trainable} and question answering~\cite{molla2007named}. A variety of methods have been proposed to tackle this task. In early approaches, manual engineering is greatly employed to design domain-specific features and heuristic rules~\cite{nadeau2007survey}. Moreover, several probabilistic models, including \textit{Hidden Markov Models (HMM)}~\cite{bikel1998nymble}, \textit{Maximum Entropy Models (ME)}~\cite{borthwick1998description}, and \textit{Conditional Random Fields (CRF)}~\cite{mccallum2003early} have been employed within supervised learning frameworks to perform this task. More recently, deep learning techniques have become dominant solutions to the NER task and have achieved promising results~\cite{li2020survey}. Neural NER models leverage deep learning architectures such as \textit{Long Short-Term Memory (LSTM)}~\cite{greff2016lstm}, \textit{Convolutional neural network (CNN)}~\cite{alzubaidi2021review}, and \textit{Bidirectional Encoder Representations from Transformers (BERT)}~\cite{devlin2018bert}, to learn contextual representations of tokens. To generate output label sequences, they use various decoding strategies, including CRFs, softmax classifiers, and span-based methods. Besides, language models can be fine-tuned with additional structures built on top of them to address specific downstream tasks. For example, the NER task can be formulated as a \textit{Machine Reading Comprehension (MRC)} task, which can be solved by fine-tuning a BERT model~\cite{sun2021biomedical}. Building on this paradigm, our study fine‑tunes a CodeBERT backbone by appending a lightweight token‑classification layer, enabling end‑to‑end NER without introducing additional task‑specific modules.

\subsection{BERT and Its Variants}
\textit{Bidirectional Encoder Representations from Transformers (BERT)}~\cite{devlin2018bert} is a transformer-based architecture that has demonstrated a strong ability to model long-range dependencies in natural language by leveraging bidirectional attention mechanisms. This capacity enables BERT to effectively capture contextual information, making it highly suitable for a wide range of natural language understanding tasks. Following the success of BERT, numerous \textit{Pre-trained Transformer-based Models (PTMs)} have been proposed, many of which have achieved state-of-the-art performance across diverse NLP benchmarks~\cite{jin2020bert}.

General-purpose BERT-based PTMs may face limitations in certain scenarios due to factors such as large model size, scalability issues, or suboptimal training strategies. To address these challenges, a variety of BERT-based variants have been proposed, each introducing modifications to the original BERT architecture or training setup to better suit specific requirements. For example, RoBERTa~\cite{liu2019roberta} retains BERT’s core architecture but improves performance by removing the next sentence prediction task, dynamically changing masking patterns during training, and increasing the size of the training corpus and training duration. ALBERT~\cite{lan2019albert} enhances BERT's scalability by introducing parameter reduction techniques such as factorized embedding parameterization and cross-layer parameter sharing, which significantly reduce the number of model parameters without sacrificing accuracy. DistilBERT~\cite{sanh2019distilbert} follows a different strategy by applying knowledge distillation to compress BERT into a lighter model. It reduces the number of parameters by approximately 40\% while retaining over 95\% of BERT’s language understanding capabilities, making it more suitable for resource-constrained environments.

While these general-purpose models work well in a broad range of NLP tasks, they often struggle with domain-specific texts such as those found in software engineering, which are rich in technical jargon and structured expressions that differ significantly from standard natural language. Therefore, researchers have adapted the BERT architecture to handle program-related data, such as source code and technical discussions, and proposed various variants of BERT models as well. CodeBERT~\cite{feng2020codebert} is trained on paired data consisting of source code and corresponding natural language descriptions, enabling it to generate semantically rich embeddings for both types of input. This dual understanding allows CodeBERT to perform well in various downstream tasks such as code search~\cite{feng2020codebert}, code summarization~\cite{feng2020codebert}, automated program repair~\cite{mashhadi2021applying}, and question answering~\cite{huang2021cosqa}. Similarly, BERTOverflow~\cite{tabassum2020code} is fine-tuned on data from Stack Overflow. It incorporates named entity recognition to better handle technical terms and identifiers commonly used in developer discussions, thereby improving its performance on software-related tasks.

\section{Experiment Setup}
\label{sec:experiment}
In this section, we begin by introducing the sensitive attributes found in software logs, then describe the datasets used for training our model and outline our annotation methodology. Finally, we present the design of our model in identifying sensitive attributes in logs.

\subsection{Sensitivite Attributes in Software Logs}
\label{sec:sensitive}
Software logs contain various attributes, including IP addresses, timestamps, and log levels. As discussed in Section~\ref{sec:related_work_pii}, this study adopts the definition of sensitivity in software logs proposed by Aghili et al.~\cite{aghili2025protecting}, classifying IP addresses, MAC addresses, host names, file paths, IDs, URLs, usernames, port numbers, and configuration details as commonly recognized sensitive attributes in software logs.
Table~\ref{tab:sensitive_attributes_examples} presents these attributes along with an example for each.

\subsection{The Benchmark Dataset}
We use the log datasets from LogHub~\cite{zhu2023loghub} to build and test our sensitive attribute detection framework. LogHub contains 16 datasets collected from both open-source and commercial systems across different domains of \textit{distributed systems} (HDFS, Hadoop, Spark, Zookeeper, OpenStack), \textit{supercomputers} (BGL, HPC, Thunderbird), \textit{operating systems} (Windows, Linux, Mac), \textit{mobile systems} (Android, HealthApp), \textit{server applications} (Apache, OpenSSH), and \textit{standalone software} (Proxifier). These datasets have been widely adopted in prior research on software logs~\cite{jiang2024large, jiang2024lilac, shan2024face}. Each dataset includes a subset of 2,000 logs with manually extracted log templates serving as the ground truth. Throughout the remainder of this paper, we refer to this collection of subsets (16 datasets with 2,000 logs each, totaling 32,000 logs) as the Benchmark Dataset.

\begin{table*}[t]
\caption{Types and examples of sensitive attributes in software logs}
\small
\begin{tabular} {ll}
\toprule
\textbf{Attribute} & \textbf{Example} \\
\toprule

IP Address & \textit{Invalid user webmaster from \textcolor{red}{173.234.31.186}} \\ 
MAC address & \textit{ARPT: 621131.293163: wl0: Roamed or switched channel, reason \#8, bssid \textcolor{red}{5c:50:15:4c:18:13}, last RSSI -64} \\
Host name & \textit{\textcolor{red}{proxy.cse.cuhk.edu.hk}: 5070 close, 0 bytes sent, 0 bytes received, lifetime 00:01}\\
File path & \textit{workerEnv.init() ok \textcolor{red}{/etc/httpd/conf/workers2.properties}} \\
ID & \textit{Verification succeeded for \textcolor{red}{blk\_-4980916519894289629}} \\
URL & \textit{the url = \textcolor{red}{http://baike.baidu.com/item/\%E8\%93\%9D\%E9\%87\%87\%E5\%92\%8C/462624?fr=aladdin}} \\
Username & \textit{Invalid user \textcolor{red}{webmaster} from 173.234.31.186}\\
Port number & \textit{proxy.cse.cuhk.edu.hk: \textcolor{red}{5070} close, 0 bytes sent, 0 bytes received, lifetime 00:01}\\
Configuration details & \textit{mapResourceRequest:<\textcolor{red}{memory:1024}, \textcolor{red}{vCores:1}>} \\

\bottomrule
\end{tabular}
\label{tab:sensitive_attributes_examples}
\end{table*}

\subsection{Sensitive Attribute Annotation Process}
\label{sec:annotation}
Our sensitive attribute detection framework relies on fine-tuning a pre-trained \textit{large language model (LLM)}. Annotated training data is required to fine-tune the model in order to identify sensitive attributes in software logs. We use the Benchmark Dataset, labeling each word in each of the 32,000 log lines with its corresponding attribute. The annotation follows the \textit{IOB (Inside-Outside-Beginning)} format~\cite{ramshaw1999text}, where \textit{B-} denotes the beginning of an attribute, \textit{I-} is used for subsequent words in the attribute, and \textit{O} indicates non-attribute words. Each sensitive attribute is annotated with its abbreviation along with the respective IOB prefix, while non-sensitive words are marked as \textit{O}.


To prepare the data, we preprocess each log message by splitting it into tokens using whitespace as the delimiter, treating each resulting word as a token for sequence labeling. This approach works well for most attributes, such as URLs, file paths, and IDs. However, network-related information, including IP addresses, host names, and port numbers, often appear concatenated (e.g., \textit{10.250.18.114:50010}), with no whitespace separating them. Standard tokenization fails to isolate such components. To handle this issue, we introduce a new category, \textit{net}, as a parent attribute of IP addresses, host names, and port numbers. This design choice enables the model to recognize and classify network-related entities even when they are concatenated, ensuring accurate detection despite the absence of whitespace separation. 

\begin{figure}[t]
\centering
\includegraphics[width=0.8\textwidth]{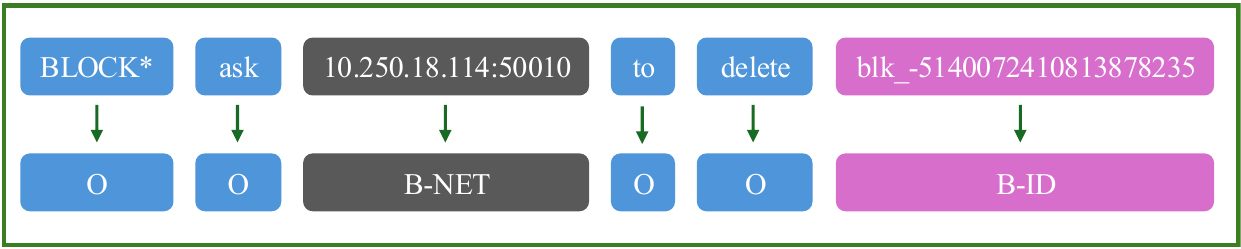}
\caption{An example of our data annotation process}
\label{fig:annotation}    
\end{figure}

Figure~\ref{fig:annotation} demonstrates the annotation process for a log message from the \textit{HDFS} dataset. In the example, the log message is \textit{BLOCK* ask 10.250.18.114:50010 to delete  blk\_-5140072410813878235}. The words \textit{BLOCK*}, \textit{ask}, \textit{to}, and \textit{delete}, are non-sensitive words, so we annotate them as \textit{O}. The word \textit{10.250.18.114:50010} is a combination of IP address and port number, therefore it belongs to the network category and we annotate it as \textit{B-net}; and finally, we annotate the \textit{blk\_-5140072410813878235} as \textit{B-ID}.

To further address the challenge of detecting concatenated network-related entities, we derive a specialized subset of the Benchmark Dataset containing only net attributes, and we refer to it as the Net Benchmark Dataset. In this variant, we preprocess network-related tokens by explicitly separating their components using special characters. For example, a combined network token such as \textit{10.250.18.114:50010} is split into two distinct tokens of \textit{10.250.18.114} and \textit{50010}. This allows us to assign more specific labels, annotating them as \textit{B-IP} and \textit{B-PORT}, respectively. This separation enables the model to better distinguish between different types of network information, improving its ability to accurately detect and classify sensitive net attributes.

In order to annotate the sensitive attributes in the Benchmark Dataset, the first two authors of this study (hereafter referred to as coders) independently performed an open coding procedure on 200 log templates randomly selected from all the log templates of all the datasets. The annotation methodology follows the systematic guidelines proposed by Basili et al.~\cite{basili1986experimentation} and is similar to prior studies~\cite{li2023did, chen2021demystifying}.

\textit{Step 1: First Round of Coding, Discussion, and Revision.}
The Benchmark Dataset contains 1,363 log templates in total. To initiate the annotation process, we randomly sampled 100 templates. Each coder independently annotated the sensitive attributes. During this phase, the coders did not have a shared annotation protocol. After completing their individual annotations, the coders shared their responses and discussed their annotation reasoning. The goal of this discussion session was to resolve inconsistencies and establish a shared understanding. Finally, each coder revised their initial annotations according to the consensus reached.

\textit{Step 2: Measuring Reliability for First Round.}
Establishing reliability is crucial to validate the annotation process~\cite{artstein2008inter}. We evaluated inter-coder agreement using both Cohen’s kappa~\cite{cohen1960coefficient} and accuracy, as Cohen’s kappa alone can underestimate agreement in datasets like ours, where most tokens are labeled as non-sensitive (i.e., annotated as \textit{O}). In the first round, the coders achieved a Cohen’s kappa of 0.53 and an accuracy of 0.82, with most disagreements related to the annotation of IDs and configuration details.

\textit{Step 3: Second Round of Coding, Discussion, and Revision.}
Following the initial round, the coders conducted another cycle of independent coding, discussion, and revision on a new random sample of 100 log templates. By this stage, a clearer and more consistent understanding of the annotation guidelines had been developed.

\textit{Step 4: Measuring Reliability for Second Round.} After the second round, we again measured the inter-coder agreement. The Cohen’s kappa improved significantly to 0.87, with an accuracy of 0.94. A Cohen’s kappa greater than 0.80 indicates strong reliability in the annotation process~\cite{mchugh2012interrater}, confirming that a consistent annotation guideline was successfully established.

\textit{Step 5: Full Dataset Annotation.} Upon achieving a high level of agreement and establishing a clear annotation protocol, the remaining log templates in the Benchmark Dataset were annotated by the first coder.

Table~\ref{tab:sensitive_attributes_percentage} presents the detailed statistics of the Benchmark Dataset. Among the 16 datasets, \textit{Apache} contains the fewest log templates with only 6 templates, while \textit{Mac} has the most with 341 templates. The median number of templates across datasets is 50. The proportion of sensitive attributes also varies significantly; \textit{HealthApp} contains no sensitive attributes, whereas every template in \textit{HDFS} and \textit{Proxifier} includes sensitive attributes. Overall, based on the median across all datasets, approximately one-third of the log templates (32.4\%) contain sensitive attributes. Among the different types of sensitive information, net (i.e., IP addresses, host names, and port numbers) is the most common, appearing in 11.2\% of log templates in median. It is followed by ID and file path attributes, with median percentages of 9.8\% and 5.6\%, respectively. The remaining attributes (i.e., MAC addresses, URLs, usernames, and configuration details) have a median percentage of 0, indicating that less than half of the datasets contain these attributes. For instance, although 48.1\% of \textit{OpenSSH} templates contain usernames, 13 out of the 16 datasets have no usernames at all. These observations highlight the need for a generalizable sensitive attribute detection pipeline capable of handling various attributes across diverse datasets, as relying on a fixed set of attributes could result in missing important sensitive information.

\begin{table*}[t]
\caption{Proportion of sensitive attributes in the Benchmark Dataset}
\small
\resizebox{1\textwidth}{!}{
\begin{tabular} {l|lllllllll}
\toprule
\textbf{Dataset} & \textbf{\# Templates} & \textbf{Sensitive (\%)} & \textbf{net (\%)} & \textbf{MAC (\%)} & \textbf{File path (\%)} & \textbf{ID (\%)} & \textbf{URL (\%)} & \textbf{Username (\%)} & \textbf{Config (\%)} \\
\hline

    Android & 166 & 10.2 & 0.0 & 0.0 & 0.6 & 9.6 & 0.0 & 0.0 & 0.0 \\

    Apache & 6 & 66.7 & 16.7 & 0.0 & 33.3 & 33.3 & 0.0 & 0.0 & 0.0 \\

    BGL & 120 & 22.5 & 6.7 & 3.3 & 12.5 & 3.3 & 0.0 & 0.0 & 0.0 \\

    Hadoop & 114 & 58.8 & 14.0 & 0.0 & 5.3 & 42.1 & 1.8 & 0.0 & 5.3 \\

    HDFS & 14 & 100.0 & 71.4 & 0.0 & 14.3 & 100.0 & 0.0 & 0.0 & 0.0 \\

    HealthApp & 75 & 0.0 & 0.0 & 0.0 & 0.0 & 0.0 & 0.0 & 0.0 & 0.0 \\

    HPC & 46 & 26.1 & 15.2 & 0.0 & 0.0 & 21.7 & 0.0 & 0.0 & 0.0 \\

    Linux & 118 & 28.8 & 7.6 & 0.0 & 2.5 & 6.8 & 0.0 & 2.5 & 15.3 \\

    Mac & 341 & 18.2 & 5.3 & 0.6 & 7.3 & 1.8 & 4.4 & 0.0 & 0.6 \\

    OpenSSH & 27 & 88.9 & 70.4 & 0.0 & 0.0 & 22.2 & 0.0 & 48.1 & 0.0 \\

    OpenStack & 43 & 95.3 & 16.3 & 0.0 & 27.9 & 65.1 & 0.0 & 0.0 & 16.3 \\

    Proxifier & 8 & 100.0 & 100.0 & 0.0 & 0.0 & 0.0 & 0.0 & 0.0 & 0.0 \\

    Spark & 36 & 47.2 & 8.3 & 0.0 & 2.8 & 25.0 & 11.1 & 0.0 & 11.1 \\

    Thunderbird & 149 & 26.8 & 6.0 & 4.0 & 6.0 & 2.7 & 0.0 & 6.0 & 9.4 \\

    Windows & 50 & 20.0 & 0.0 & 0.0 & 10.0 & 10.0 & 0.0 & 0.0 & 2.0 \\

    Zookeeper & 50 & 36.0 & 22.0 & 0.0 & 6.0 & 8.0 & 0.0 & 0.0 & 0.0 \\ \hline

    \textbf{Median} & \textbf{50} & \textbf{32.4} & \textbf{11.2} & \textbf{0.0} & \textbf{5.6} & \textbf{9.8} & \textbf{0.0} & \textbf{0.0} & \textbf{0.0} \\

    \bottomrule
 \end{tabular}
    }
\label{tab:sensitive_attributes_percentage}
\end{table*}

\subsection{SDLog: Sensitivity Detector in Software Logs}
We propose SDLog, a deep learning-based framework designed to detect sensitive attributes in software logs. SDLog can determine whether a word in a log message corresponds to sensitive information and, if so, assign it to the appropriate category (e.g., username). Researchers and developers can leverage SDLog as an alternative to regular expression-based approaches for identifying sensitive data that needs to be anonymized.

We formulate the detection task as a sequence tagging problem, a well-established approach in the field of \textit{Natural Language Processing (NLP)}~\cite{collobert2011natural, huang2015bidirectional}. Sequence tagging is commonly used for \textit{Named Entity Recognition (NER)}; for example, to identify people, organizations, and locations in a sentence. Similarly, in our context, given a log message, SDLog classifies each word as either non-sensitive or sensitive, and labels sensitive words with their corresponding categories.

To build SDLog, we leverage CodeBERT, a pre-trained large language model released by Microsoft~\cite{feng2020codebert}. CodeBERT extends the BERT architecture to handle both natural language and programming languages. It is a bimodal model trained jointly on paired natural language and code data. Specifically, it was trained on six popular programming languages: Python, Java, JavaScript, PHP, Ruby, and Go, using a combination of masked language modeling and replaced token detection objectives. CodeBERT consists of 12 Transformer encoding layers with a total of 125 million parameters. It supports a variety of downstream tasks, including code-to-code translation, code-to-text generation, text-to-code synthesis, and text-to-text tasks. Additionally, it has demonstrated strong performance on tasks such as code search and automatic code documentation generation, making it a suitable model for applications involving both source code and natural language. We choose CodeBERT as the backbone of SDLog because it has been pre-trained on programming languages in addition to natural language, allowing it to better model the semi-structured, code-like patterns that are often present in software logs. Unlike models such as RoBERTa~\cite{liu2019roberta} or the original BERT~\cite{devlin2019bert}, which are exclusively pre-trained on natural language corpora, CodeBERT is specifically optimized to handle syntax, identifiers, and structural elements commonly found in code and, by extension, in log messages.

\begin{figure}[t]
\centering
\includegraphics[width=0.9\textwidth]{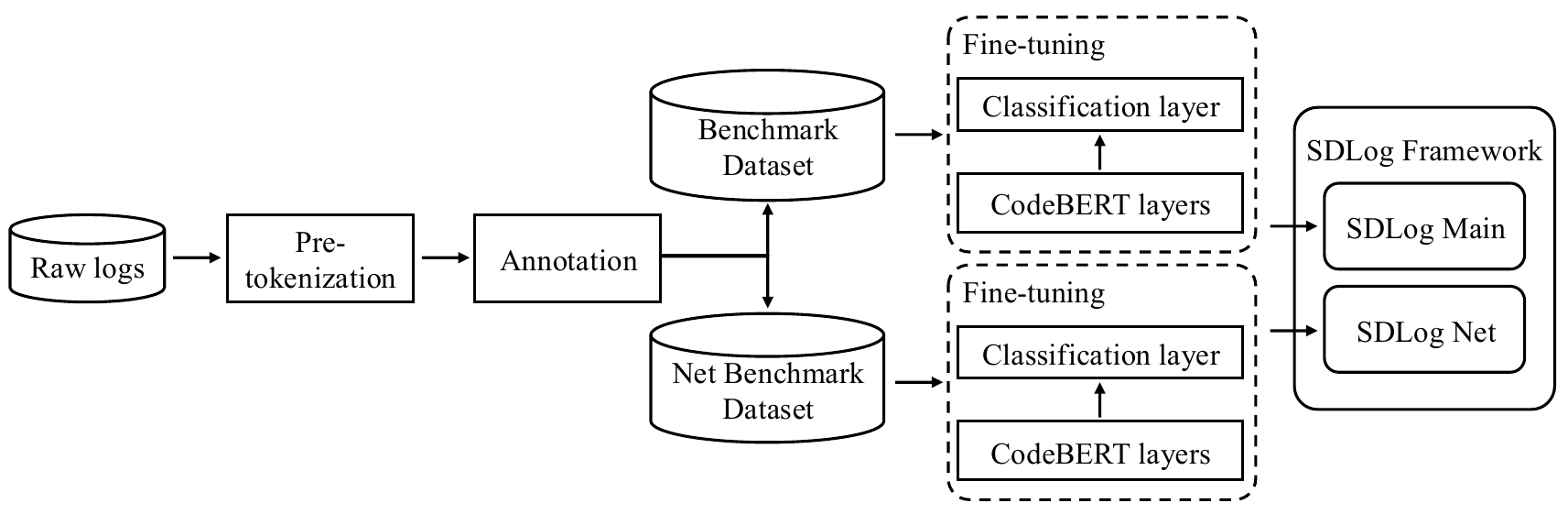}
\caption{An overall diagram of SDLog framework}
\label{fig:sdlog}    
\end{figure}

Figure~\ref{fig:sdlog} presents an overview of our approach to identify sensitive attributes in software logs. We implement SDLog as described in the following sections.

\textbf{Pre-tokenization Process.} Before applying the pre-trained tokenizer of the CodeBERT model, we follow the same procedure discussed in Section~\ref{sec:annotation}, splitting each log message using whitespace as the delimiter. As a result, each word is treated as an individual token, and these tokens are the smallest units processed by the model.

\textbf{Classification Head.} A single fully connected layer is attached to the output representations of the CodeBERT structure to project each token vector onto the IOB label space. A softmax layer then converts these mappings into probability distributions over the label classes. The entire model, including both the classifier and the underlying encoder, is fine-tuned end-to-end using a cross-entropy loss computed on the ground-truth token-level annotations.

\textbf{SDLog Hierarchy.} SDLog consists of two specialized pre-trained transformer-based models, each optimized for different levels of granularity in sensitive attribute detection. The first model, a large-scale transformer, is fine-tuned on log data annotated with eight distinct categories of sensitive information: net, MAC, file path, ID, URL, username, and configuration details. This model is trained using the full Benchmark Dataset, enabling it to generalize across a broad range of attribute types and structural patterns observed in real-world logs. The second model is a lightweight transformer specifically fine-tuned to perform fine-grained classification within the broader net category. Unlike the first model, this one is trained exclusively on the Net Benchmark Dataset. It learns to distinguish between network-related entities, including IP addresses, port numbers, and host names. This hierarchical modeling approach allows SDLog to first identify coarse-grained sensitive spans and then apply a more targeted classification to resolve ambiguous network-related entities, improving precision in scenarios with complex or overlapping attribute types. Both models follow the same procedure and use the same hyperparameters.

\textbf{Hyperparameter Settings.} We fine-tune a transformer-based token classification model, initialized from the \textit{microsoft/codebert-base} checkpoint~\cite{feng2020codebert}, to perform NER over log data. The model is optimized using the AdamW optimizer with a learning rate of 5e-5 and a weight decay of 0.01. Considering the size of the log dataset, training is conducted for 2 epochs with a linear learning rate scheduler. Evaluation is performed every 300 training steps, and checkpoints are saved at the same interval. To promote generalization and avoid overfitting, the best-performing model, based on validation loss, is retained at the end of training. These hyperparameter choices follow standard practices for adapting pre-trained language models to token-level classification tasks~\citep{huggingface_token_classification}.

\textbf{Experimental Environment.} All experiments are conducted on a workstation equipped with an 11th Gen Intel Core i7-11700K CPU (8 cores, 16 threads, up to 5.0 GHz), 32 GB of RAM, and an NVIDIA GeForce GTX 1060 GPU with 3 GB of memory. The system runs a 64-bit Linux environment with Python 3.12, and all models are trained using PyTorch~\cite{paszke2019pytorch} and the Hugging Face Transformers library~\cite{wolf2019huggingface}.

\section{Research Questions and Restuls}
\label{sec:RQs}

In this section, we present our~\textit{Research Questions (RQs)} and their results. Each RQ is structured according to its motivation, approach, and findings.

\subsection{RQ1. How effective are regex-based approaches in identifying sensitive attributes in software logs?}
\label{sec:rq1}

\subsubsection{Motivation}
Identifying sensitive attributes is a crucial first step in most log anonymization pipelines, as it determines which log attributes should be anonymized. Regular expressions have traditionally been the accepted method for this task in both research and industry. However, despite their widespread adoption, there is a notable lack of standardized regex patterns even for commonly identified attributes such as IP addresses and URLs, which can lead to inconsistencies in detection. Furthermore, the effectiveness of regex-based methods across various attribute types and log formats remains largely unexplored. A systematic evaluation of these limitations can provide a clearer understanding of current practices, highlight specific weaknesses, and support the case for alternative approaches.

\subsubsection{Approach}
We perform a comprehensive search across three research databases, initially identifying 185 relevant articles. We review each article and its replication package (if available) according to inclusion and exclusion criteria to extract regex patterns that identify sensitive attributes in software logs. In addition, we collect regex patterns from three industry partners, which they use in their anonymization pipelines. Our selection process adheres to the systematic methodology proposed by Wohlin et al.~\cite{wohlin2012experimentation} and is detailed below.

\setcounter{secnumdepth}{4} 
\paragraph{Searching Research Databases}
Our literature search is conducted across three widely recognized academic databases: IEEE Xplore Digital Library\footnote{\url{https://ieeexplore.ieee.org/}}, ACM Digital Library\footnote{\url{https://dl.acm.org/}}, and Engineering Village\footnote{\url{https://www.engineeringvillage.com/}}, which itself aggregates three databases of Compendex, Inspec, and Knovel. We aim to identify conference and journal papers that use regex patterns for detecting sensitive attributes in software logs. To achieve this, we formulate the following search query:

\begin{flushleft}
\textit{(``Abstract'': log) AND (``Abstract'': privacy OR ``Abstract'': sensitiv* OR ``Abstract'': anonymiz* OR ``Abstract'': log pars*) AND (``Full Text'': regex OR ``Full Text'': ``regular expression'')}
\end{flushleft}

This query targets articles that contain the terms~\textit{log} and~\textit{regular expression} or~\textit{regex} while also falling within the domains of privacy, log anonymization, or log parsing. We include ``log pars*'' keyword because many log parsers use regex patterns as a preprocessing step~\cite{qin2024preprocessing, he2017towards}.

Applying this query, we find 133 articles from the IEEE Xplore Digital Library, 47 from the ACM Digital Library, and 38 from Engineering Village, amounting to a total of 218 articles. After removing duplicate articles, we get a total of 185 papers for further analysis.

\paragraph{Inclusion and Exclusion Criteria}
\label{sec:inclusion&exclusion}
We focus on identifying papers that use regex to detect one or more sensitive attributes discussed in Section~\ref{sec:sensitive}. We review tables, figures, and code snippets of each paper, searching for keywords such as ``regex'' and ``regular expression''. Additionally, we check whether the article includes a replication package. If available, we review it and look for signs of regex patterns, such as the presence of ``import re'' in Python projects.

We exclude articles that focus on data types other than software logs (e.g., search logs) or those that use regex patterns to identify attributes outside the scope of our defined sensitive attributes (e.g., timestamps). Additionally, we do not consider patterns that are specific to one software system. For example, Li et al.~\cite{li2022swisslog} use ``blk\_-?\textbackslash d+'' to detect block IDs in \textit{Hadoop} logs and ``core\textbackslash .\textbackslash d+'' to identify core IDs in \textit{BGL} logs. However, we do not consider these regex patterns for the ID attribute, as they are specific to particular systems.

\paragraph{Industry Regexes}
In addition to the academic literature, we also collect regex patterns from three of our industry partners. These partners are large-scale organizations, each with more than 1,000 employees, where they manage extensive log datasets derived from supercomputers, operating systems, and server application environments. The regexes provided by these partners are part of their anonymization pipelines, designed to detect sensitive attributes within their software logs prior to data sharing, either internally or externally. We apply the same inclusion and exclusion criteria discussed in Section~\ref{sec:inclusion&exclusion} to these regex patterns, capturing those that detect sensitive attributes while discarding patterns customized to specific systems or use cases (e.g., a server name pattern unique to a company).

\paragraph{Regex Extraction}
We extract regex patterns from research papers, their replication packages, and industry partners, organizing them by attribute type (e.g., IP address and URL). Each article, replication package, or industry partner may provide regex patterns for one or more log attributes.

\paragraph{Evaluating Regex Patterns}
To systematically evaluate the effectiveness of the extracted regex patterns, we run each regex pattern individually against the Benchmark Dataset, which includes 32,000 logs from 16 different datasets. For each execution, we compute standard evaluation metrics including precision, recall, and F1-score, based on token-level ground truth annotations. True positives are counted when the regex-matched text overlaps with the labeled sensitive attributes; false positives are counted when a match occurs outside that sensitive attribute region; and false negatives capture labeled sensitive attributes that were not detected using the regex pattern. Therefore, in this evaluation, precision measures how many of the matches found by the regex patterns were correct, and recall measures how many of the true labeled attributes the regex patterns successfully found. This analysis enables a detailed comparison of regex patterns, highlighting their strengths and weaknesses in identifying different types of structured information within heterogeneous logs.

\subsubsection{Restuls} \hspace*{-\leftskip}

\begin{table}[p]
\centering
\small
\caption{Performance of individual regular expressions in categorizing sensitive attributes on the Benchmark Dataset}
\begin{tabular}{l|l|l|l|l|l}
\toprule
& \textbf{IP address} & \textbf{P (\%)} & \textbf{R (\%)} & \textbf{F1 (\%)} & \textbf{Source} \\ \hline
1 & \texttt{(/|)([0-9]+\textbackslash .)\{3\}[0-9]+(:[0-9]+|)(:|)} & \textbf{\textbf{92.1}} & 85.1 & 88.4 &~\cite{dai2023pilar, he2017drain, yu2023brain, xu2023hue} \\
2 & \texttt{([0-9.]+)\textbackslash s} & 11.4 & 48.8 & 18.5 &~\cite{zhang2022research} \\
3 & \texttt{([0-9]+.)\{3\}[0-9]+(:[0-9]+)} & 23.0 & 0.6 & 1.1 &~\cite{huang2020paddy} \\
4 & \texttt{((\textbackslash d+).(\textbackslash d+).(\textbackslash d+).(\textbackslash d+))} & 32.5 & \textbf{99.7} & 49.0 &~\cite{yu2024unlocking} \\
5 & \texttt{(\textbackslash d)+3\textbackslash d(:\textbackslash d+)?} & 8.6 & 9.6 & 9.1 &~\cite{niu2023fsmflog} \\
6 & \texttt{(\textbackslash d+\textbackslash .)\{3\}\textbackslash d+(:\textbackslash d+)?} & \textbf{92.1} & 85.1 & 88.4 &~\cite{messaoudi2018search} \\
7 & \texttt{\textasciicircum(25[0-5]|2[0-4]\textbackslash d|[0-1]?\textbackslash d?\textbackslash d)(\textbackslash .(25[0-5]|2[0-4]\textbackslash d|[0-1]?\textbackslash d?\textbackslash d))\{3\}\$} & 0.0 & 0.0 & 0.0 &~\cite{naumiuk2014anonymization} \\
8 & \texttt{(\textbackslash b\textbackslash d\{1,3\}(?:\textbackslash .\textbackslash d\{1,3\})\{3\}\textbackslash b)} & \textbf{92.1} & 85.1 & \textbf{88.5} &~\cite{wang2024vcrlog} \\
9 & \texttt{(\textbackslash d\{1,3\}(?:\textbackslash .\textbackslash d\{1,3\})\{3\}):?\textbackslash d*} & \textbf{92.1} & 85.1 & \textbf{88.5} &~\cite{wang2024vcrlog} \\
10 & \texttt{\textbackslash d+\textbackslash .\textbackslash d+\textbackslash .\textbackslash d+\textbackslash .\textbackslash d+} & \textbf{92.1} & 85.1 & 88.4 &~\cite{li2022swisslog} \\
11 & \texttt{(\textbackslash d\{1,3\}\textbackslash .\textbackslash d\{1,3\}\textbackslash .\textbackslash d\{1,3\}\textbackslash .\textbackslash d\{1,3\})[,: )]} & 90.7 & 78.6 & 84.2 &~\cite{li2023glad} \\
12 & \texttt{[0-9]\{1,3\}.[0-9]\{1,3\}.[0-9]\{1,3\}.[0-9]\{1,3\}} & 31.5 & \textbf{99.7} & 47.9 &~\cite{debnath2018loglens}\\
13 & \texttt{(/|)(\textbackslash d+.)\{3\}\textbackslash d+(:\textbackslash d+)?} & 32.5 & \textbf{99.7} & 49.1 &~\cite{qin2024preprocessing} \\
14 & \texttt{[0-9]+\textbackslash .[0-9\textbackslash .:]*[0-9]} & 56.7 & 85.1 & 68.1 &~\cite{li2024revisiting} \\
15 & \texttt{(\textbackslash d\{1,3\}\textbackslash .\textbackslash d\{1,3\}\textbackslash .\textbackslash d\{1,3\}\textbackslash .\textbackslash d\{1,3\})} & \textbf{92.1} & 85.1 & \textbf{88.5} & Company 1 \\
16 & \texttt{\textbackslash b\textbackslash d\{1,3\}(?:\textbackslash .\textbackslash d\{1,3\})\{2,\}\textbackslash b} & 91.8 & 85.1 & 88.3 & Company 2 \\
17 & \texttt{(\textbackslash b\textbackslash d\{1,3\}\textbackslash .)(\textbackslash d\{1,3\}\textbackslash .)(\textbackslash d\{1,3\}\textbackslash .)(\textbackslash d\{1,3\}\textbackslash b)} & \textbf{92.1} & 85.1 & \textbf{88.5} & Company 3 \\ \bottomrule

& \textbf{MAC address} & \textbf{P (\%)} & \textbf{R (\%)} & \textbf{F1 (\%)} & \textbf{Source} \\ \hline \
1 & \texttt{\textasciicircum([0-9A-Fa-f]\{2\}[:-])\{5\}([0-9A-Fa-f]\{2\})\$} & 0.0 & 0.0 & 0.0 &~\cite{qin2024preprocessing} \\ \bottomrule

& \textbf{File path} & \textbf{P (\%)} & \textbf{R (\%)} & \textbf{F1 (\%)} & \textbf{Source} \\ \hline
1 & \texttt{((((?<!\textbackslash w)[A-Z,a-z]:)|(\textbackslash .\{1,2\}\textbackslash \textbackslash))([\textasciicircum \textbackslash b\%\textbackslash /\textbackslash |:\textbackslash n\textbackslash "]*))|(\textbackslash "\textbackslash 2([\textasciicircum \%\textbackslash /\textbackslash |:\textbackslash n\textbackslash "]*)\textbackslash ")|} & 59.0 & 98.3 & 73.7 &~\cite{li2023glad} \\ 

& \texttt{((?<!\textbackslash w)(\textbackslash .\{1,2\})?(?<!\textbackslash /)(\textbackslash /((\textbackslash \textbackslash \textbackslash b)|[\textasciicircum \textbackslash b\%\textbackslash |:\textbackslash n\textbackslash "\textbackslash \textbackslash \textbackslash /])+)+\textbackslash /?)} & & & & \\ 

2 & \texttt{/[\textbackslash w/. :-]+} & 55.6 & 99.5 & 71.3 &~\cite{wahab2024secret} \\

3 & \texttt{(/[\textasciicircum /\textbackslash s]+)+} & 48.1 & 99.5 & 64.9 &~\cite{wahab2024secret} \\ 

4 & \texttt{(([A-Z]:)|)(/\textbackslash S+)+} & 47.8 & 99.5 & 64.5 &~\cite{xu2023hue} \\ 

5 & \texttt{(/|)(([\textbackslash w.-]+|\textbackslash <\textbackslash *\textbackslash >)/)+([\textbackslash w.-]+|\textbackslash <\textbackslash *\textbackslash >)} & \textbf{66.7} & 98.1 & \textbf{79.4} &~\cite{qin2024preprocessing} \\ 

6 & \texttt{([A-Za-z]:|\textbackslash .)\{0,1\}(/|\textbackslash \textbackslash )[0-9A-Za-z\textbackslash -\_\textbackslash .:/\textbackslash *\textbackslash +\textbackslash \$\#@!\textbackslash \textbackslash \textbackslash ?=\%\&]+(?<![:\textbackslash .])} & 48.1 & \textbf{100.0} & 65.0 &~\cite{li2024revisiting} \\

7 & \texttt{\textbackslash /(\textbackslash S+)} & 47.8 & 99.5 & 64.5 & Company 3\\ \bottomrule

& \textbf{ID} & \textbf{P (\%)} & \textbf{R (\%)} & \textbf{F1 (\%)} & \textbf{Source} \\ \hline
1 & \texttt{(?:UUID|GUID|version|id)[\textbackslash \textbackslash =:\textbackslash "\textbackslash '\textbackslash s]*\textbackslash b[a-fA-F0-9]\{8\}-[a-fA-F0-9]\{4\}-[a-fA-F0-9]} & 0.0 & 0.0 & 0.0 &~\cite{wahab2024secret} \\

& \texttt{\{4\}-[a-fA-F0-9]\{4\}-[a-fA-F0-9]\{12\}\textbackslash b} & & & & \\

2 & \texttt{<([\textasciicircum >]+)>} & 2.6 & 0.1 & 0.2 &~\cite{wahab2024secret} \\ 

3 & \texttt{[pP]id[:|-|=|\textbackslash s/]*(\textbackslash d+)} & 97.7 & 1.3 & 2.5 &~\cite{li2023glad} \\

4 & \texttt{[uU]id[:|-|=|\textbackslash s/]*(\textbackslash d+)} & \textbf{99.8} & \textbf{23.5} & \textbf{38.1} &~\cite{li2023glad} \\ \bottomrule

& \textbf{URL} & \textbf{P (\%)} & \textbf{R (\%)} & \textbf{F1 (\%)} & \textbf{Source} \\ \hline
1 & \texttt{[A-Za-z\textbackslash .]+://[A-Za-z0-9\textbackslash .\textbackslash /\textbackslash +\#@:\_\textbackslash -]+(?<![:\textbackslash .])} & 91.4 & 99.2 & \textbf{95.1} &~\cite{li2024revisiting} \\

2 & \texttt{(https?://\textbackslash S+)} & \textbf{100.0} & 39.1 & 56.2 &~\cite{li2023glad} \\

3 & \texttt{https?://[\textasciicircum \textbackslash s\#]+\#[A-Za-z0-9\textbackslash -\textbackslash =\textbackslash +]+} & 0.0 & 0.0 & 0.0 &~\cite{wahab2024secret} \\

4 & \texttt{http[s]?://(?:[a-zA-Z]|[0-9]|[\$-\_@.\&+]|[!*\textbackslash \textbackslash (\textbackslash \textbackslash).]|(?:\textbackslash \%[0-9afA-F][0-9a-fA-F]))+} & \textbf{100.0} & 39.1 & 56.2 &~\cite{wahab2024secret} \\

5 & \texttt{([\textbackslash w-]+\textbackslash .)+[\textbackslash w-]+(:\textbackslash d+)?} & 0.9 & \textbf{100.0} & 1.8 &~\cite{yu2023log} \\

6 & \texttt{(\textbackslash S+\textbackslash .\textbackslash S+(\textbackslash .\textbackslash S+)+(:\textbackslash d+)?)|(\textbackslash w+-\textbackslash w+(-\textbackslash w+)+)} & 0.7 & \textbf{100.0} & 1.4 &~\cite{xu2023hue} \\

7 & \texttt{\textbackslash bhttps?://(www.)?[a-zA-Z0-9-]+(.[a-zA-Z]\{2,\})+(:[0-9]\{1,5\})?(/[\textasciicircum \textbackslash s]*)?\textbackslash b} & \textbf{100.0} & 31.2 & 47.6 &~\cite{qin2024preprocessing} \\ \bottomrule

& \textbf{Username} & \textbf{P (\%)} & \textbf{R (\%)} & \textbf{F1 (\%)} & \textbf{Source} \\ \hline
1 & \texttt{user( |  )[A-Za-z0-9]+(?<!request)(?! methods)} & \textbf{42.0} & 25.3 & 31.6 &~\cite{li2024revisiting} \\
2 & \texttt{user\textbackslash :\textbackslash s(\textbackslash w+)} & 0.0 & 0.0 & 0.0 &~\cite{yadav2023identification} \\
3 & \texttt{r?[uU]ser[:|-|=|\textbackslash s/]*<(\textbackslash w+)>|r?[uU]ser[:|-|=|\textbackslash s/]*(\textbackslash w+)} & 35.2 & \textbf{72.0} & \textbf{47.3} &~\cite{li2023glad} \\ \bottomrule

& \textbf{Port} & \textbf{P (\%)} & \textbf{R (\%)} & \textbf{F1 (\%)} & \textbf{Source} \\ \hline
1 & \texttt{[pP]ort[=: |:|=|: |\textbackslash s/]*(\textbackslash d{1,5})} & \textbf{96.0} & \textbf{8.1} & \textbf{15.0} &~\cite{li2023glad} \\ \bottomrule

& \textbf{Configuration details} & \textbf{P (\%)} & \textbf{R (\%)} & \textbf{F1 (\%)} & \textbf{Source} \\ \hline
1 & \texttt{size\textbackslash s+(\textbackslash d+)} & \textbf{19.2} & \textbf{14.2} & \textbf{16.3} &~\cite{wang2024vcrlog} \\ 
\bottomrule
\end{tabular}
\label{tab:regex_rq1}
\end{table}

\textbf{No common regex ground truth exists to identify log sensitive attributes.}
Table~\ref{tab:regex_rq1} summarizes the complete set of regex patterns identified through our systematic search. In total, we collected 41 distinct regex patterns across 8 different attribute types, namely IP addresses, MAC addresses, file paths, IDs, URLs, usernames, ports, and configuration details. We did not find any regex patterns for host names. Among the found regexes, 17 patterns specifically target IP addresses and are used by 20 different sources. Similarly, for other attributes such as file paths, IDs, URLs, and usernames, each source appears to have developed its own set of regex patterns. This leads to our first key observation: there is no unified or widely adopted set of regex patterns that researchers and practitioners rely on. Instead, every article, project, or company that requires the identification of sensitive attributes within software logs seems to independently curate and maintain its own collection of regex patterns.

\textbf{Regex patterns show large performance variability for the same attribute.}
Our evaluation reveals that regex patterns designed for the same attribute can exhibit vastly different levels of performance depending on subtle variations in their construction. For instance, among the regexes for IP addresses, F1-scores range from extremely low values such as 0.0\%, 1.1\%, and 9.1\%, to relatively high values such as 88.4\% and 88.5\%. Similarly, for regexes designed to detect URLs, we observe F1-scores as low as 0.0\%, 1.4\%, and 1.8\%, alongside patterns achieving F1-scores up to 95.1\%. These results reveal two major challenges in designing and using regex patterns: every small detail in the pattern matters, and regex patterns often lack generalizability across different log formats and datasets.

\textbf{Small differences in regex design have a large impact.}
To illustrate the importance of fine details, consider two regex patterns intended to detect IP addresses: (1) \texttt{((\textbackslash d+).(\textbackslash d+).(\textbackslash d+).(\textbackslash d+))} and (2) \texttt{\textbackslash d+\textbackslash .\textbackslash d+\textbackslash .\textbackslash d+\textbackslash .\textbackslash d+}. Both patterns look similar, as each aims to match four numeric blocks separated by dots. However, their performance differs considerably. The first pattern achieves a very high recall of 99.7\%, successfully detecting almost all IP addresses present in the dataset. Nevertheless, its precision is low, 32.5\%, indicating that it also incorrectly identifies many non-IP strings as IP addresses. In contrast, the second pattern yields a high precision of 92.1\%, correctly labeling true IP addresses in most cases, but has a slightly lower recall of 85.0\%. Consequently, the first pattern reaches an F1-score of 49.0\%, while the second achieves a much higher F1-score of 88.4\%. This example shows how small differences in regex structure can significantly impact precision, recall, and overall effectiveness.

\textbf{Regex patterns often lack generalizability.}
Generalizability presents another critical limitation. For example, the regex pattern \texttt{\textasciicircum([0-9A-Fa-f]\{2\}[:-])\{5\}([0-9A-Fa-f]\{2\})\$} completely failed to detect any MAC addresses in the Benchmark Dataset. This failure is due to the pattern's assumption that the MAC address must be the only content in the log message. In real-world logs, however, attributes are typically embedded among various other strings. With a small modification to the pattern and changing it to \texttt{\textbackslash b([0-9A-Fa-f]\{2\}[:-])\{5\}([0-9A-Fa-f]\{2\})\textbackslash b}, we observe a dramatic improvement, with a precision of 98.8\%, recall of 100.0\%, and F1-score of 99.4\%. A similar issue was observed with an IP address pattern \texttt{\textasciicircum(25[0-5]|2[0-4]\textbackslash d|[0-1]?\textbackslash d?\textbackslash d)(\textbackslash .(25[0-5]|2[0-4]\textbackslash d|[0-1]?\textbackslash d?\textbackslash d))\{3\}\$}. By changing it to \texttt{\textbackslash b(25[0-5]|2[0-4]\textbackslash d|[0-1]?\textbackslash d?\textbackslash d)(\textbackslash .(25[0-5]|2[0-4]\textbackslash d|[0-1]?\textbackslash d?\textbackslash d))\{3\}\textbackslash b}, the initial F1-score of 0.0\% increases to 88.5\%. These findings emphasize that regex patterns often fail to generalize across different log structures and datasets unless they are explicitly designed to handle common variations in log formatting.

\textbf{Regex patterns can achieve reasonable results in certain attributes that follow common patterns (e.g., IP addresses or file paths.}
Looking at the best-performing regex patterns in Table~\ref{tab:regex_rq1}, shown in bold, we find that regexes are capable of achieving reasonable results for specific types of sensitive attributes, particularly IP addresses, MAC addresses (with the improved pattern discussed in the last paragraph), file paths, and URLs. The best F1-score in these attributes are 88.5\%, 99.4\%, 79.4\%, and 95.1\%, respectively. This relatively high performance is largely due to the fact that these attributes often follow strict and predictable patterns that remain consistent across a variety of log formats and datasets. Nevertheless, it is important to highlight that these figures reflect the best-performing regex patterns. The overall performance across all regexes for a given attribute shows more variability; for example, the median F1-score for file paths is 65.0\% and for URLs is 47.6\%, indicating that many regex patterns still struggle to generalize effectively even within structured attribute types.

\textbf{However, regex patterns are not well-suited for detecting certain attributes that are constrained to rigid structures (e.g., IDs or usernames).}
As shown in Table~\ref{tab:regex_rq1}, regex patterns generally perform poorly for detecting certain attributes such as IDs, usernames, ports, and configuration details. This is primarily due to the absence of a consistent and rigid structure for these attributes, which makes them difficult for regex patterns to capture effectively. The data reflects this, with fewer researchers and practitioners developing regex patterns for these attributes. For example, while we identified 17 distinct regex patterns for detecting IP addresses, only 4, 3, 1, and 1 patterns were found for IDs, usernames, ports, and configuration details, respectively. We also did not find any regex patterns for host names. The lack of consistent patterns for these attributes is widely recognized, and as a result, fewer efforts are made to address them with regex. Hence, these attributes are often overlooked in detection efforts, with anonymization pipelines typically focusing on more structured attributes such as IP addresses and file paths. For those who do attempt to use regex for these attributes, the results are often of low quality. For instance, the best-performing regex pattern for detecting IDs achieved an F1-score of just 38.1\%, while the best pattern for detecting usernames had an F1-score of 47.3\%. The performance for detecting ports and configuration details is even worse, with F1-scores of 15.0\% and 16.3\%, respectively. These attributes often appear in a variety of formats, making them even harder to detect. For example, usernames may or may not be preceded by the keyword \textit{user}, complicating the task for regex patterns designed with specific assumptions. Similarly, ports can be located in various positions within a log; immediately after an IP address, immediately after a host name, or embedded in text, making it a challenging task for strict regex patterns to distinguish them from other integers.

\newenvironment{mybox}[1]{%
    \begin{tcolorbox}[title={Summary of RQ1}]%
    }{
    Our analysis shows that there is no common ground truth for regex patterns to detect sensitive attributes in software logs; instead, researchers and practitioners tend to develop their own patterns, leading to inconsistency and fragmentation. Regex performance varies significantly even within the same attribute type, with small design differences causing large impacts on precision and recall. While regex can achieve reasonable results for well-structured attributes such as IP addresses, it performs poorly for less structured attributes such as IDs, usernames, ports, and configuration details. Overall, regex patterns often lack generalizability across datasets and are not well-suited for attributes without strict formatting.
    \end{tcolorbox}
}
\begin{mybox}{}
\end{mybox}

\subsection{RQ2. How effective is SDLog in identifying sensitive attributes in software logs?}
\label{sec:rq2}

\subsubsection{Motivation}
Although regular expressions provide a structured, rule-based method for detecting sensitive information in software logs, they often fall short when applied to the wide range of log formats and datasets, limiting their practical utility. In contrast, machine learning methods, such as SDLog, have the potential to address these shortcomings by leveraging contextual and structural information within the data. To demonstrate the effectiveness of SDLog, We evaluate its ability to accurately detect sensitive attributes in previously unseen log datasets. Because log formats vary significantly across systems in terms of syntax, terminology, and structure, testing the model's ability to generalize is essential. By comparing SDLog to the best-performing regex patterns, we establish a performance baseline and assess whether a learning-based solution is suitable for deployment in dynamic, heterogeneous logging environments.

\subsubsection{Approach}
We evaluate the effectiveness of SDLog in identifying sensitive attributes and compare its results with the best-performing regex patterns found in RQ1.

\paragraph{Baseline}
As our baseline, we select the best-performing regex pattern for each attribute from Table~\ref{tab:regex_rq1}, based on the highest F1-score. In cases where multiple regex patterns for the same attribute achieve identical F1-scores, we randomly choose one. This results in a set of eight regex patterns, each corresponding to one of the following attributes: IP address, MAC address, file path, ID, URL, username, port, and configuration details. Using these patterns, we design an anonymization pipeline that simulates real-world usage. In contrast to our evaluation on RQ1 where we ran each regex pattern separately in order to find the best regex patterns, in this RQ and to build our regex baseline, each log line is passed through all eight regexes. When a pattern detects a match, the matched value is replaced with a placeholder (e.g., an IP match is replaced with {\it\$IP}). 

Prior research has suggested that the order in which regex patterns are applied can impact the outcome~\cite{qin2024preprocessing}. For example, if a loosely written ID regex partially matches an IP address and is executed before the IP regex, the true IP may be incorrectly masked as {\it\$ID}. However, if the IP regex is executed first, it correctly identifies and replaces the IP address. To evaluate the impact of regex execution order, we design a sampling strategy. Out of the 8! possible permutations of the eight regex patterns, we randomly sample 500 orders and run each through our anonymization pipeline. Finally, we report the minimum and maximum precision, recall, and F1-score values observed for each sensitive attribute across these different orderings.

\paragraph{SDLog Evaluation}
We use cross-validation to evaluate the performance of SDLog. Given the variability in syntax and structure across software logs, this step is critical for evaluating the model’s performance on unseen data. For the first, large-scale model of SDLog, we use \textit{Leave-One-Out Cross-Validation (LOOCV)}~\cite{wong2015performance}, training the model on all datasets except one and testing it on the excluded dataset. This procedure is repeated until each dataset has been used as the test set once. By conducting LOOCV over our diverse set of logs, we obtain a rigorous and comprehensive evaluation of SDLog’s effectiveness in identifying sensitive attributes across varying log formats and sources. For the second model focused on net sub-attribute classification, we use 4-fold cross-validation~\cite{fushiki2011estimation}, splitting the 16 datasets into 12 for training and 4 for testing in each round. 4-fold cross validation is more suitable than LOOCV for the second model, as three datasets contain no net attributes, making k-fold validation a better fit for balanced evaluation.

We evaluate SDLog through a two-step process. First, we measure its ability to detect sensitive attributes in datasets it has not encountered during training. These attributes correspond to the nine categories outlined in Section~\ref{sec:sensitive}. The goal is to determine whether SDLog can accurately locate these sensitive attributes. Second, we evaluate SDLog’s ability to classify the detected attributes into their respective sensitive categories. This two-tiered evaluation ensures that SDLog not only generalizes well to new data but also effectively distinguishes between different types of sensitive data.

\subsubsection{Restuls} \hspace*{-\leftskip}

\begin{table*}[t]
\caption{Performance of regular expression pipeline in categorizing sensitive attributes on the Benchmark Dataset with randomized orders (500 runs)}
\small
\begin{tabular} {l|l|ll|ll|ll}
\toprule
& & \multicolumn{2}{c}{\textbf{Percision (\%)}} & \multicolumn{2}{c}{\textbf{Recall (\%)}} & \multicolumn{2}{c}{\textbf{F1 (\%)}} \\ \cline{3-8} 
                   \textbf{Attribute} & \textbf{Best Regex Pattern} & \textbf{Min} & \textbf{Max} & \textbf{Min} & \textbf{Max} & \textbf{Min} &  \textbf{Max}  \\
                   
\hline

    IP & \texttt{(\textbackslash b\textbackslash d\{1,3\}(?:\textbackslash .\textbackslash d\{1,3\})\{3\}\textbackslash b)} & 92.1 & 93.0 & 85.1 & 85.1 & 88.5 & 88.9  \\
    MAC & \texttt{\textbackslash([0-9A-Fa-f]\{2\}[:-])\{5\}([0-9A-Fa-f]\{2\})\textbackslash} & 98.6 & 98.6 & 100.0 & 100.0 & 99.3 & 99.3 \\
    File path & \texttt{(/|)(([\textbackslash w.-]+|\textbackslash <\textbackslash *\textbackslash >)/)+([\textbackslash w.-]+|\textbackslash <\textbackslash *\textbackslash >)} & 66.6 & 68.4 & 97.7 & 98.0 & 79.3 & 80.5 \\
    ID & \texttt{[uU]id[:|-|=|\textbackslash s/]*(\textbackslash d+)} & 99.8 & 99.8 & 23.5 & 23.5 & 38.0 & 38.0 \\
    URL & \texttt{[A-Za-z\textbackslash .]+://[A-Za-z0-9\textbackslash .\textbackslash /\textbackslash +\#@:\_\textbackslash -]+(?<![:\textbackslash .])} & 70.6 & 94.6 & 9.4 & 99.2 & 16.6 & 95.1 \\
    Username & \texttt{r?[uU]ser[:|-|=|\textbackslash s/]*<(\textbackslash w+)>|r?[uU]ser[:|-|=|\textbackslash s/]*(\textbackslash w+)} & 36.6 & 36.6 & 72.0 & 72.0 & 48.5 & 48.5 \\
    Port & \texttt{[pP]ort[=: |:|=|: |\textbackslash s/]*(\textbackslash d{1,5})} & 96.0 & 96.2 & 8.1 & 8.1 & 15.0 & 15.0 \\
    Configuration & \texttt{size\textbackslash s+(\textbackslash d+)} & 19.2 & 19.2 & 14.2 & 14.2 & 16.3 & 16.3 \\

    \bottomrule
 \end{tabular}
\label{tab:rq2_regex}
\end{table*}

\textbf{The order of regex patterns significantly affects performance.}  
Table~\ref{tab:rq2_regex} presents the results of applying our top-performing regex patterns for detecting sensitive attributes in software logs, executed 500 times with randomized pattern orders. Despite the distinct syntactic structures among the eight sensitive categories, the performance of regex-based sensitive attribute detection varies notably based on pattern ordering. In some cases, the variation is minor; for example, the F1-score for file paths ranges from 79.3\% to 80.5\%, and for IP addresses from 88.5\% to 88.9\%. However, for URL pattern, the F1-score fluctuates drastically, from as low as 16.6\% up to 95.1\%, representing a 78.5\% difference. This highlight three key observations: (1) the sequence in which regex patterns are applied has a substantial impact on overall performance, (2) the optimal ordering is likely dataset-dependent and must be tuned manually through trial and error, and (3) regex-based approaches lack robustness and reliability due to their sensitivity to manual configurations, raising concerns about their scalability and trustworthiness in real-world log datasets.

\textbf{SDLog can accurately detect sensitive information in software logs, achieving an overall F1-score of 92.9\%.}
Table~\ref{tab:rq2_model_sensitivity} summarizes SDLog’s performance on the Benchmark Dataset, reporting precision (i.e., the proportion of correctly predicted sensitive attributes among all predictions), recall (i.e., the proportion of actual sensitive attributes correctly identified), F1-score (i.e., the harmonic average of precision and recall), and support (i.e., the number of predictions per dataset). The support values vary considerably across datasets, depending on how many sensitive attributes they contain. For example, the \textit{HealthApp} dataset contains no actual sensitive information (see Table~\ref{tab:sensitive_attributes_percentage}), yet the model falsely predicted one instance as sensitive. In contrast, SDLog achieved perfect F1-scores of 100\% on datasets such as \textit{Apache} and \textit{Proxifier}, where it successfully identified all sensitive attributes. To compute the overall performance, we aggregate true positives, false positives, and false negatives across all testing datasets from the cross-validation to build a unified confusion matrix. Overall, SDLog achieved a precision of 94.6\%, recall of 91.2\%, and F1-score of 92.9\%.

\textbf{SDLog demonstrates strong performance in categorizing most sensitive attribute types.}
Looking at Table~\ref{tab:rq2_SDLog_main_attributes}, SDLog achieves high precision, recall, and F1-scores for the majority of sensitive categories. For example, it correctly detects 97.8\% of net attributes and is accurate 99.5\% of the time. Similarly, it identifies 94.8\% of file paths with a precision of 99.9\%. For more challenging attributes, such as ID, username, and configuration details, which lack consistent structure and vary significantly across datasets, SDLog still performs well, achieving F1-scores of 88.7\%, 84.8\%, and 50.4\% respectively. Although the model achieves strong overall performance, its lower scores on MAC and URL attributes can be primarily explained by the limited availability of labeled examples: only 70 instances of MAC and 128 instances of URL are present across the entire Benchmark Dataset, with even fewer examples available during training. Moreover, these two attributes appear in just three datasets (check Table~\ref{tab:sensitive_attributes_percentage}), and their values likely vary significantly between training and testing sets. This scarcity and inconsistency make it challenging for the model to learn generalizable patterns. With more diverse labeled data, we expect SDLog’s performance on these attributes to improve substantially. In contrast, categories such as Username and net have significantly more training examples (e.g., 1,623 and 13,851 respectively), enabling the model to learn their patterns more effectively. Despite these limitations, SDLog proves to be a highly effective tool for detecting and categorizing sensitive attributes in logs. Future work can address lower-performing categories by incorporating additional labeled data and targeted fine-tuning, further enhancing SDLog’s coverage and robustness.

\begin{table*}[t]
\caption{SDLog performance in detecting sensitive attributes across the Benchmark Dataset}
\small
\begin{tabular} {l|llll}
\toprule
\textbf{Dataset} & \textbf{Percision (\%)} & \textbf{Recall (\%)} & \textbf{F1 (\%)} & \textbf{Support} \\
\hline

    Android & 80.5 & 89.9 & 84.9 & 313 \\
    Apache & 100.0 & 100.0 & 100.0 & 1481 \\
    BGL & 100.0 & 86.3 & 92.6 & 175 \\
    Hadoop & 98.8 & 80.0 & 88.4 & 2082 \\
    HDFS & 93.1 & 95.1 & 94.1 & 4417 \\
    HealthApp & 0.0 & 0.0 & 0.0 & 1 \\
    HPC & 100.0 & 68.8 & 81.5 & 369 \\
    Linux & 99.8 & 99.7 & 99.7 & 3874 \\
    Mac & 62.3 & 49.9 & 55.4 & 577 \\
    OpenSSH & 92.0 & 91.6 & 91.8 & 5363 \\
    OpenStack & 100.0 & 88.8 & 94.1 & 3559 \\
    Proxifier & 100.0 & 100.0 & 100.0 & 3042 \\
    Spark & 62.3 & 64.2 & 63.2 & 2162 \\
    Thunderbird & 97.1 & 86.5 & 91.5 & 980 \\
    Windows & 99.8 & 99.0 & 99.4 & 1207 \\
    Zookeeper & 99.9 & 99.8 & 99.8 & 1271 \\ \hline
    \textbf{Overall} & \textbf{94.6} & \textbf{91.2} & \textbf{92.9} & \textbf{30873} \\

    \bottomrule
 \end{tabular}
\label{tab:rq2_model_sensitivity}
\end{table*}

\textbf{SDLog’s second-level granularity achieves near-perfect categorization of net sub-attributes.}
As presented in Table~\ref{tab:rq2_SDLog_net_attribute}, the second, lightweight model of SDLog is dedicated to distinguishing between the sub-attributes of the net attribute, namely, IP addresses, port numbers, and host names. Trained with a high number of support for each attribute, SDLog achieves high F1-scores of 99.7\% for IP addresses, 100.0\% for Port numbers, and 99.6\% for host names. This two-stage approach, first detecting net-related entities, then categorizing the sub-attributes of it, proves effective in handling attributes that frequently appear in similar context and overlapping or compound forms (e.g., IP:port or host:port). 

\textbf{SDLog significantly outperforms regular expressions in detecting sensitive attributes in software logs.} 
A comparison between SDLog (Tables~\ref{tab:rq2_SDLog_main_attributes} and~\ref{tab:rq2_SDLog_net_attribute}) and the regex-based approach (Table~\ref{tab:rq2_regex}) reveals that SDLog achieves higher accuracy in identifying and classifying sensitive attributes. Please note that in Table~\ref{tab:rq2_regex}, the~\textit{Max} values represent the highest performance achieved by the regular expression patterns, based on the optimal ordering and selection of the best-performing pattern for each attribute. Even under these optimal conditions, regex can detect IP addresses with 85.1\% recall and 93.0\% precision. In contrast, SDLog detects 97.8\% of net attributes with 99.5\% precision and further classifies IP addresses with 99.4\% recall and 100.0\% precision in the second granularity level, resulting in over a 10\% improvement in F1-score. This result is especially important, as IP addresses are widely regarded as the most sensitive attribute in software logs, recognized as sensitive by academia, industry, and privacy regulations~\cite{aghili2024understanding}.

For file paths, regex performs well in recall (98.0\%) but poorly in precision (68.4\%), yielding a relatively modest F1-score. SDLog, by contrast, balances both dimensions effectively, achieving 99.9\% precision and 94.8\% recall, improving the F1-score by 16.8\%. Port number detection is another area where regex underperforms, with a recall of just 8.1\% and F1-score of 15.0\%. On the other hand, SDLog first achieves an F1-score of 98.6\% on detecting the net attribute, and then, achieves an F1-score of 100.0\% on distinguishing the port number out of the net attribute. Host name detection is handled with near-perfect accuracy by SDLog, while no regex pattern was found for this attribute.

For attributes lacking a fixed format or structure, such as username, ID, and configuration details, SDLog significantly outperforms regular expressions. While regex achieves only 38.0\% F1-score for IDs, 48.5\% for usernames, and 16.3\% for configuration details, SDLog improves these scores to 88.7\%, 84.8\%, and 50.4\%, respectively, improving F1-score by 30\%-50\%. These results demonstrate that SDLog not only outperforms regex across all categories, but particularly excels where rule-based methods are most limited.

\begin{table*}[t]
\caption{SDLog performance in categorizing sensitive attributes on the Benchmark Dataset}
\small
\begin{tabular} {l|llll}
\toprule
\textbf{Attribute} & \textbf{Percision (\%)} & \textbf{Recall (\%)} & \textbf{F1 (\%)} & \textbf{Support} \\
\hline

    Net & 99.5 & 97.8 & 98.6 & 13851 \\
    MAC & 100.0 & 40.0 & 57.1 & 70 \\
    File path & 99.9 & 94.8 & 97.3 & 2868 \\
    ID & 86.0 & 91.5 & 88.7 & 9745 \\
    URL & 0.0 & 0.0 & 0.0 & 128 \\
    Username & 99.8 & 73.7 & 84.8 & 1623 \\
    Configuration & 95.7 & 34.2 & 50.4 & 1049 \\

    \bottomrule
 \end{tabular}
\label{tab:rq2_SDLog_main_attributes}
\end{table*}

\begin{table*}[t]
\caption{SDLog performance in categorizing net attribute on the Net Benchmark Dataset}
\small
\begin{tabular} {l|llll}
\toprule
\textbf{Attribute} & \textbf{Percision (\%)} & \textbf{Recall (\%)} & \textbf{F1 (\%)} & \textbf{Support} \\
\hline

    IP & 100.0 & 99.4 & 99.7 & 8922 \\
    Port & 100.0 & 100.0 & 100.0 & 7168 \\
    Host name & 100.0 & 99.2 & 99.6 & 6013 \\

    \bottomrule
 \end{tabular}
\label{tab:rq2_SDLog_net_attribute}
\end{table*}

\newenvironment{mybox2}[1]{%
    \begin{tcolorbox}[title={Summary of RQ2}]%
    }{
    Our results demonstrate that SDLog outperforms regex-based approaches in detecting sensitive information in software logs. Regex performance is highly sensitive to the ordering of patterns, making it unreliable and difficult to scale, particularly for attributes with non-structured formats. In contrast, SDLog achieves robust and accurate sensitive attribute detection across diverse datasets, with an overall F1-score of 92.9\%. SDLog especially excels in handling attributes without predefined structure, such as username, ID, and configuration details, highlighting its effectiveness and generalization capability in real-world log datasets.
    \end{tcolorbox}
}
\begin{mybox2}{}
\end{mybox2}

\subsection{RQ3. How effective is SDLog in identifying sensitive attributes when fine-tuned on a target dataset?}
\label{sec:rq3}

\subsubsection{Motivation}
While RQ2 shows that SDLog outperforms regular expressions, especially in complex or unstructured attributes, its performance can still be limited for certain categories that lack sufficient labeled examples or structural diversity in training. Our goal, however, is to develop a sensitive attribute detection pipeline that can be practically adopted by organizations, researchers, and practitioners as a reliable alternative to the commonly used regular expressions in anonymization workflows. In this research question, we explore how fine-tuning SDLog with a subset of logs from the target dataset can improve its sensitive attribute detection performance. We evaluate the impact of different amounts of fine-tuning data on both overall sensitive attribute detection performance and attribute-level categorization, using the same evaluation procedure as in RQ2.

\begin{table*}[t]
\caption{Performance of fine-tuned SDLog for individual target datasets (with 20, 50, and 100 logs) in detecting sensitive attributes across the Benchmark Dataset}
\small
\begin{tabular} {l|lll|lll|lll}
\toprule
& \multicolumn{3}{c}{\textbf{20}} & \multicolumn{3}{c}{\textbf{50}} & \multicolumn{3}{c}{\textbf{100}} \\ \cline{2-10}
                   \textbf{Dataset} & \textbf{P (\%)} & \textbf{R (\%)} & \textbf{F1 (\%)} & \textbf{P (\%)} & \textbf{R (\%)} & \textbf{F1 (\%)} & \textbf{P (\%)} & \textbf{R (\%)} & \textbf{F1 (\%)} \\

\hline

    Android & 92.8 & 99.0 & 95.8 & 88.0 & 99.0 & 93.6 & 86.7 & 100.0 & 92.9 \\
    Apache & 100.0 & 100.0 & 100.0 & 100.0 & 100.0 & 100.0 & 100.0 & 100.0 & 100.0 \\
    BGL & 85.7 & 92.3 & 88.9 & 98.5 & 100.0 & 99.2 & 85.5 & 100.0 & 92.2 \\
    Hadoop & 80.6 & 84.5 & 82.5 & 81.4 & 92.5 & 86.8 & 82.8 & 99.6 & 90.4 \\
    HDFS & 100.0 & 100.0 & 100.0 & 100.0 & 100.0 & 100.0 & 100.0 & 100.0 & 100.0 \\
    HPC & 97.1 & 98.1 & 97.6 & 54.3 & 98.1 & 69.9 & 96.3 & 100.0 & 98.1 \\
    Linux & 99.9 & 99.6 & 99.8 & 99.5 & 99.6 & 99.6 & 99.6 & 99.6 & 99.6 \\
    Mac & 67.0 & 58.5 & 62.5 & 68.3 & 85.6 & 76.0 & 91.1 & 96.2 & 93.6 \\
    OpenSSH & 100.0 & 99.7 & 99.8 & 100.0 & 99.8 & 99.9 & 100.0 & 99.8 & 99.9 \\
    OpenStack & 100.0 & 88.7 & 94.0 & 100.0 & 95.0 & 97.4 & 100.0 & 99.8 & 99.9 \\
    Proxifier & 100.0 & 100.0 & 100.0 & 100.0 & 100.0 & 100.0 & 100.0 & 100.0 & 100.0 \\
    Spark & 82.6 & 97.9 & 89.6 & 99.7 & 97.9 & 98.8 & 97.0 & 97.9 & 97.5 \\
    Thunderbird & 96.9 & 87.7 & 92.1 & 98.1 & 87.6 & 92.6 & 92.4 & 95.3 & 93.8 \\
    Windows & 100.0 & 99.2 & 99.6 & 100.0 & 99.2 & 99.6 & 100.0 & 99.2 & 99.6 \\
    Zookeeper & 85.4 & 100.0 & 92.1 & 91.7 & 99.9 & 95.6 & 88.8 & 99.9 & 94.0 \\ \hline
    \textbf{Overall} & \textbf{96.2} & \textbf{96.3} & \textbf{96.3} & \textbf{96.9} & \textbf{98.1} & \textbf{97.5} & \textbf{97.4} & \textbf{99.5} & \textbf{98.4} \\

    \bottomrule
 \end{tabular}
\label{tab:rq3_model_sensitivity}
\end{table*}

\subsubsection{Approach}
To evaluate how fine-tuning improves SDLog's ability to detect sensitive attributes in a new dataset, we simulate a practical scenario where only a small subset of labeled log samples from the target dataset is available for adaptation. Specifically, we consider three fine-tuning sizes: 20, 50, and 100 logs. For this purpose, we begin with the base models trained during RQ2 (i.e., models trained on all datasets except the target one), and further fine-tune them using a small, labeled subset from the target dataset.

For each dataset in our Benchmark Dataset, which contains 2,000 logs, we reserve the first 100 logs that contain at least one sensitive attribute as the fine-tuning set. The remaining 1,900 logs are used as the fixed test set. For example, in RQ2, to evaluate SDLog on the \textit{Apache} dataset, we trained the model on the other 15 datasets and tested it on all 2,000 \textit{Apache} logs. In this research question, we extend that setup by additionally fine-tuning the pre-trained SDLog model using 20, 50, or 100 \textit{Apache} logs before testing it, each time using the same fixed 1,900-log test set for consistency.

This fine-tuning and evaluation process is repeated for each dataset in the Benchmark Dataset. At the end of this process, we report the impact of different fine-tuning sizes (20, 50, and 100 logs) on the model’s performance. We evaluate both its ability to detect the presence of sensitive information and its accuracy in categorizing the specific type of each sensitive attribute. For this RQ, we exclude the \textit{HealthApp} dataset from our evaluation setup, as it does not contain any sensitive attributes in the entire dataset, and therefore, it cannot be used for fine-tuning SDLog.

\subsubsection{Restuls} \hspace*{-\leftskip}

\textbf{Fine-tuning SDLog with target-specific data enhances its sensitive attribute detection performance.} 
Table~\ref{tab:rq3_model_sensitivity} summarizes the results of sensitive attribute detection across the Benchmark Dataset when SDLog is fine-tuned using 20, 50, and 100 labeled log samples from the target dataset. In general, increasing the number of fine-tuning examples leads to improved performance. Specifically, in 13 out of 15 datasets, fine-tuning with 100 logs results in equal or higher F1-scores compared to using only 20 logs. Notably, when fine-tuned with 100 samples, SDLog achieves an F1-score above 90\% and a recall above 95\% on all datasets. This indicates that, across the entire Benchmark Dataset, SDLog, when fine-tuned with just 100 logs, can accurately detect at least 95\% of all sensitive attributes. Overall, using 100 fine-tuning logs, SDLog achieves an average precision of 97.4\% (up by 1.2\% compared to using 20 logs), a recall of 99.5\% (up by 3.2\% compared to using 20 logs), and an F1-score of 98.4\% (a 2.1\% improvement compared to using 20 logs).

\textbf{Fine-tuning SDLog with target-specific logs significantly improves the categorization of sensitive attributes.} 
Table~\ref{tab:rq3_SDLog_main_attributes} shows the categorization performance of SDLog after fine-tuning with varying amounts of labeled data from the target dataset. The results show that fine-tuning greatly enhances the model’s ability to distinguish between different types of sensitive attributes. As the number of fine-tuning examples increases, from 20 to 100 logs, SDLog demonstrates consistent improvement gains across all attribute types. Even for attributes that already had high performance using 20 logs as fine-tuning, such as net, file path, ID, and username, fine-tuning with 100 logs yields additional improvements in F1-score ranging from 0.4\% to 2.3\%. More substantial gains are observed in challenging attributes. For instance, comparing the 20 log samples and 100 log samples for fine-tuning, the F1-score for URL increases from 76.9\% to 93.4\% (a 16.5\% improvement), for MAC address from 54.5\% to 95.9\% (a 41.4\% boost), and for configuration details, the most irregular and unstructured category, from 45.4\% to 95.1\% (a 49.7\% increase). Notably, with just 20 fine-tuning logs, 4 out of 7 attributes already surpass a 90\% F1-score, and when using 100 logs, all 7 attributes exceed 90\% F1-score, highlighting the strong impact of even modest fine-tuning on SDLog’s classification capabilities.

\begin{table*}[t]
\caption{Performance of fine-tuned SDLog for individual target datasets (with 20, 50, and 100 logs) in categorizing sensitive attributes on the Benchmark Dataset}
\small
\begin{tabular} {l|lll|lll|lll}
\toprule
& \multicolumn{3}{c}{\textbf{20}} & \multicolumn{3}{c}{\textbf{50}} & \multicolumn{3}{c}{\textbf{100}} \\ \cline{2-10} 
                   \textbf{Attribute} & \textbf{P (\%)} & \textbf{R (\%)} & \textbf{F1 (\%)} & \textbf{P (\%)} & \textbf{R (\%)} & \textbf{F1 (\%)} &  \textbf{P (\%)} & \textbf{R (\%)} & \textbf{F1 (\%)} \\
\hline

    Net & 96.5 & 99.5 & 98.0 & 96.1 & 99.8 & 97.9 & 96.9 & 99.9 & 98.4 \\
    MAC & 94.7 & 38.3 & 54.5 & 100.0 & 74.5 & 85.4 & 92.2 & 100.0 & 95.9 \\
    File path & 99.1 & 95.9 & 97.5 & 99.4 & 98.3 & 98.8 & 99.4 & 98.9 & 99.1 \\
    ID & 95.4 & 99.4 & 97.4 & 96.8 & 99.8 & 98.3 & 97.6 & 99.9 & 98.7 \\
    URL & 100.0 & 62.5 & 76.9 & 100.0 & 62.5 & 76.9 & 99.0 & 88.4 & 93.4 \\
    Username & 94.2 & 98.0 & 96.1 & 99.8 & 98.0 & 98.9 & 97.5 & 99.4 & 98.4 \\
    Configuration & 97.2 & 29.6 & 45.4 & 96.7 & 62.7 & 76.1 & 96.5 & 93.7 & 95.1 \\

    \bottomrule
 \end{tabular}
\label{tab:rq3_SDLog_main_attributes}
\end{table*}

\newenvironment{mybox3}[1]{%
    \begin{tcolorbox}[title={Summary of RQ3}]%
    }{
    Our results show that fine-tuning SDLog on a small, labeled subset of the target dataset significantly improves its performance in both detecting and categorizing sensitive attributes. With as few as 100 fine-tuning samples, SDLog accurately identifies 99.5\% of sensitive attributes in Benchmark Dataset and achieve an F1-score of 98.4\%. Fine-tuning also enhances the model's ability to classify specific attribute types, with special improvements for more challenging categories such as configuration details. When fine-tuned with 100 samples, SDLog reaches an F1-score exceeding 90\% across all categories of sensitive attributes. These findings highlight the practical value of lightweight fine-tuning in adapting SDLog to new environments and improving its effectiveness.
    \end{tcolorbox}
}
\begin{mybox3}{}
\end{mybox3}

\section{Discussion}
\label{sec:discussion}
\subsection{When are regular expressions suitable for detecting sensitive attributes in software logs?}
Combining results from our first two research questions, we identify several inherent limitations of regular expressions for detecting sensitive attributes in software logs. First, there is no standardized ground truth for regex patterns; for example, we found 17 distinct patterns for identifying IP addresses, a relatively well-defined format. Second, performance varies widely across patterns; some regexes for IP detection gained extremely low F1-scores (e.g., 0.0\%, 1.1\%, and 9.1\%). Third, even small syntactic differences between similar-looking regexes can lead to large performance variations. Fourth, regexes often lack generalizability across datasets, requiring developers to craft dataset-specific patterns, which limits reuse and scalability. Fifth, regexes perform poorly on attributes that have no defined format and are context-dependent, such as username, ID, and configuration details. These attributes are often ignored or weakly captured in practice. Finally, in real-world pipelines where multiple regexes are applied sequentially, the overall performance is highly sensitive to their execution order; an order that must be optimized per dataset but is not known in advance.

Despite these limitations, regexes can still be effective in certain controlled scenarios. Our results show that when the most effective patterns are carefully selected and applied in an optimal sequence, regexes can perform well on attributes with clearly defined structures, such as MAC addresses or URLs, achieving F1-scores above 95\%. However, for more ambiguous attributes such as username, configuration details, or ID, regexes remain unreliable even under ideal conditions. We also did not find any usable regex patterns for detecting host names.

\subsection{What are the advantages of using a learning-based model such as SDLog for detecting sensitive attributes in software logs?}
Adopting a deep learning model such as SDLog offers several key advantages over traditional regex-based methods. Unlike regexes, SDLog does not rely on hand-crafted patterns and can identify a wide range of sensitive attributes, including those that lack rigid or well-defined formats (e.g., usernames, IDs, and configuration details). Additionally, SDLog eliminates the problem of dependencies on the order of rules. Since it uses a unified model for all attribute types, it avoids conflicts and interferences that often arise when multiple regexes are applied sequentially. One of SDLog’s key advantages is its generalizability: it can serve as a pre-trained model for sensitive attribute detection in software logs, eliminating the need for developers to manually design and validate regex patterns for every new environment. Furthermore, SDLog benefits from contextual understanding, allowing it to detect sensitive information based not just on format, but also on surrounding semantics, something regex patterns are inherently incapable of. In summary, SDLog offers a more robust, flexible, and scalable solution for sensitive data detection in software logs.

\subsection{How easy is it to integrate SDLog into real-world log anonymization pipelines?}
As shown in the results of RQ2, even the plain SDLog model outperforms the best-performing regular expressions. Integrating SDLog into an anonymization pipeline is straightforward. SDLog, trained on all 16 datasets of the Benchmark Dataset, is publicly available on Hugging Face (see Section~\ref{sec:data}), and can be used like any standard pre-trained language model such as BERT. After formatting the logs appropriately, developers can run SDLog with a simple API call. Moreover, RQ3 shows that fine-tuning SDLog on as few as 20 to 100 labeled log lines from the target dataset significantly boosts performance. Based on our experience, labeling 100 logs typically takes only 2–3 days, and fine-tuning the model takes about a few minutes on a standard PC with a consumer-grade graphics card. To support adoption, we also provide the scripts needed to prepare the raw dataset and fine-tune the SDLog. By releasing our models and tools, we aim to make SDLog an accessible and practical alternative to regex-based methods in real-world anonymization pipelines.

\section{Threats to Validity}
\label{sec:threats}

\hspace*{\parindent}\textbf{Internal validity.}
The definition of \textit{sensitive} information may vary across organizations, individuals, and legal jurisdictions. To mitigate this, we adopted the definition proposed by Aghili et al.~\cite{aghili2024understanding}, which combines perspectives on sensitivity from academic literature, industrial practice, and privacy regulations. 

\textbf{External validity.}
Although we evaluated SDLog on 16 publicly available log datasets that have been widely used in prior research on software logs~\cite{jiang2024large, jiang2024lilac, shan2024face}, the results may not generalize to all logging environments, such as those used in enterprise or government environments. Nonetheless, we demonstrated that fine-tuning SDLog on a small portion of a new dataset leads to substantial performance gains, suggesting the method is adaptable to unseen domains.

Additionally, SDLog may be less effective at detecting sensitive attributes that were not present in the training data, such as geolocation coordinates or API keys. However, our fine-tuning experiments show that even attributes that were underrepresented during pre-training can be accurately detected after domain-specific fine-tuning.

\textbf{Construct validity.}
The ground truth annotations were created manually by the first two authors. They may include subjective judgment regarding what constitutes sensitive information. To mitigate potential bias, we followed a systematic annotation approach and measured inter-rater agreement, achieving a high reliability after two rounds of independent coding followed by discussion sessions.

Another potential threat lies in the selection of regex patterns, which may not reflect the most effective rules an experienced practitioner could design. To mitigate this, we systematically collected patterns from three academic databases and integrated regexes from three industry collaborators. We found a total of 41 regex patterns across eight sensitive attribute types and benchmarked only the top-performing patterns in our comparisons.

There is also a risk of overfitting, particularly when SDLog is fine-tuned on small datasets, potentially causing it to learn dataset-specific artifacts. We addressed this by employing an early stopping strategy during training to prevent overfitting.

\section{Conclusion}
\label{sec:conclusion}
We introduce SDLog, a deep learning-based framework for detecting sensitive attributes in software logs, a critical first step for log anonymization. While traditional approaches rely heavily on manually designed regular expressions, these methods suffer from limited generalizability and poor performance on unstructured or context-dependent data. SDLog overcomes these limitations by leveraging contextual understanding through pre-trained language models, enabling it to generalize across diverse log formats and structures. In our evaluation across 16 datasets, SDLog consistently outperforms the best-performing regex patterns in identifying sensitive information. When fine-tuned with as few as 100 samples from a target dataset, SDLog achieves near-perfect detection, correctly identifying 99.5\% of sensitive attributes. Our approach could be integrated into real-world log anonymization pipelines conveniently. Future work can further enhance SDLog’s capabilities by incorporating additional log datasets and increasing coverage for underrepresented attribute types.

\section{Data Availability}
\label{sec:data}
To facilitate the integration of SDLog into log anonymization pipelines, we release our models, datasets, and supporting scripts. Both the SDLog Main and SDLog Net models are hosted on Hugging Face:~\url{https://huggingface.co/LogSensitiveResearcher/SDLog_main} and~\url{https://huggingface.co/LogSensitiveResearcher/SDLog_net}.
In addition, we provide the annotated dataset of 32,000 log lines used to train SDLog, along with scripts for dataset preparation and fine-tuning. These resources are available on our GitHub repository:~\url{https://github.com/mooselab/SDLog}.

\section*{Acknowledgments}
We would like to gratefully thank the Natural Sciences and Engineering Research Council of Canada (NSERC, \#RGPIN-2021-03900) and the Department of Computer and Software Engineering at Polytechnique Montreal (BSFD funds) for their funding support for this work.

\bibliographystyle{ACM-Reference-Format}
\bibliography{acmart}

\end{document}